\documentclass{aa}  

\usepackage{graphicx}
\usepackage{hyperref}
\usepackage{txfonts}
\begin{document} 

   \title{Galaxy populations in the Hydra I cluster from the VEGAS survey}
   \subtitle{III. The realm of low surface brightness features and intra-cluster light}

   \author{Marilena Spavone\inst{1} \fnmsep
   \thanks{\email{marilena.spavone@inaf.it}}
          \and Enrichetta Iodice \inst{1}
          \and Felipe S. Lohmann \inst{2}
          \and Magda Arnaboldi \inst{2}
          \and Michael Hilker  \inst{2}
          \and Antonio La Marca\inst{1,3,9}
          \and Rosa Calvi \inst{1}
          \and Michele Cantiello \inst{4}
          \and Enrico M. Corsini \inst{5}
          \and Giuseppe D'Ago \inst{6}
          \and Duncan A. Forbes \inst{7}
          \and Marco Mirabile \inst{4,8}
          \and Marina Rejkuba \inst{2}
}
           \institute{INAF $-$ Astronomical Observatory of Capodimonte, Salita Moiariello 16, I-80131, Naples, Italy
            \and
            European Southern Observatory, Karl$-$Schwarzschild-Strasse 2, 85748 Garching bei München, Germany
            \and
            Kapteyn Institute, University of Groningen, Landleven 12, 9747 AD Groningen, the Netherlands
            \and
            INAF $-$ Astronomical Observatory of Abruzzo, Via Maggini, 64100, Teramo, Italy
            \and
            Dipartimento di Fisica e Astronomia ``G. Galilei'', Universit\`a di Padova, vicolo dell'Osservatorio 3, I-35122 Padova, Italy
            \and 
            Institute of Astronomy, University of Cambridge, Madingley Road, Cambridge CB3 0HA, UK
            \and
            Centre for Astrophysics and  Supercomputing, Swinburne University of Technology, Hawthorn VIC 3122, Australia
            \and
            Gran Sasso Science Institute, viale Francesco Crispi 7, I-67100 L'Aquila, Italy
            \and SRON Netherlands Institute for Space Research, Landleven 12, 9747 AD Groningen, The Netherlands
             }

   \date{Received 02/07/2024; accepted 26/07/2024}


  \abstract
   {In this paper, we analyse the light distribution in the Hydra I cluster of galaxies to explore their low surface brightness features, measure the intra-cluster light, and address the assembly history of the cluster. For this purpose, we used deep wide-field $g$- and $r$-band images obtained with the VLT Survey Telescope (VST) as part of the VEGAS project. The VST mosaic covers $\sim 0.4$ times the virial radius (R$_{vir}$)
   around the core of the cluster, which enabled us to map the light distribution down to faint surface brightness levels of $\mu_g \sim 28$~mag/arcsec$^2$. In this region of the cluster, 44 cluster members are brighter than $m_B \leq 16$~mag, and the region includes more than 300 dwarf galaxies. Similar to the projected distribution of all cluster members (bright galaxies and dwarfs), we find that the bulk of the galaxy light is concentrated in the cluster core, which also emits in the X-rays, and there are two overdensities: in the north (N) and south-east (SE) with respect to the cluster core. We present the analysis of the light distribution of all the bright cluster members. After removing foreground stars and other objects, we measured the diffuse intra-cluster light and compared its distribution with that of the globular clusters and dwarf galaxies in the cluster. We find that most of the diffuse light low surface brightness features, and signs of possible gravitational interaction between galaxies reside in the core and in the group in the N, while ram-pressure stripping is frequently found to affect galaxies within the SE group. All these features confirm that the mass assembly in this cluster is still ongoing. By combining the projected phase-space with these observed properties, we trace the different stages of the assembly history. We also address the main formation channels for the intra-cluster light detected in the cluster, which has a total luminosity of $L_{\rm ICL} \sim 2.2 \times 10^{11} L_{\odot}$ and contributes
   $\sim 12\%$ to the total luminosity of the cluster.
   
    }
   

   \maketitle
%

\section{Introduction}\label{sec:intro}


According to the Lambda-cold dark matter ($\Lambda$CDM) scenario, clusters of galaxies are expected 
to grow over time by accreting smaller groups
along filaments, driven by the effect of gravity that is generated by the total matter content 
\citep[e.g.][]{White-Rees1978,Bullock2001}. 
In the deep potential well at the cluster centre, the galaxies continue to undergo active mass 
assembly. In this process, gravitational interactions and merging between systems 
of comparable mass and/or smaller objects play a fundamental role in defining their morphology 
and kinematics \citep[][]{Toomre1972}. 
In particular, as a result of these events, a significant amount of debris 
is deposited at larger radii from the galaxy centre and is retained by the DM halo 
\citep[e.g.][]{Bullock2005}.
This debris can be traced by loosely bound stars and globular clusters (GCs), stellar streams, 
and tidal tails, which contribute to the build-up of the stellar haloes and the intra-cluster 
light (ICL). All these structures are fainter by more than 4 magnitudes than the central regions of 
galaxies ($\mu_g \geq 26$~mag/arcsec$^2$), they have multiple stellar populations and 
complex kinematics, and they are still growing at the present epoch. 
At larger distances from the galaxy centre, the dynamical timescales are longer,
and all these structures can therefore survive for several billion years. 
As it encodes the physical processes, the ICL is considered a highly valuable diagnostic of the mass assembly in all environments \citep{Montes2019, Contini2021, Pillepich2018, Jimenez2018, Kluge2020}.

From the theoretical side, semi-analytic models and hydrodynamic simulations give detailed 
predictions about the structure and stellar populations of stellar haloes, the ICL formation, 
and the amount of substructure in various types of environment
\citep[see][and references therein]{Cooper2015,Cook2016,Pillepich2018,Monachesi2019,Merritt2020,Contini2021,Contini2023,Tang2023}.
In particular, simulations predict that the brightest and most massive galaxies are made by the in situ component that formed at an early epoch and dominates at the central radii, as well as by the accreted ex situ component, which is assembled as a result of the gravitational interaction events \citep[][]{Cooper2013,Remus2017,Pillepich2018,Pulsoni2020,Pulsoni2021,Remus2022}. 
The ex situ component, which is identified as the stellar halo, 
extends over several kiloparsec from the galaxy centre. It appears
in a diffuse form, but can also host stellar streams, shells, or tidal tails that result 
from repeated accretion events \citep[][]{Cooper2015}. The existence, the morphology, and the stability with time of these structures depend on the mass ratio of the involved 
satellites, as studied by \citet{Mancillas2019}.

In the simulated groups and clusters of galaxies, the ICL appears as a diffuse component at 
large cluster-centric distances ($\geq 150$~kpc). It is made of unbound stars that float in the intra-
cluster space and grows with time during the mass-assembly process. The bulk of the ICL is
assembled in the redshift range $0\leq z \leq 1$, and the ICL fraction ($L_{ICL,g}/L_{tot,g}$) reaches $\sim$ 20-40$\%$
at $z = 0$ \citep[see][as reviews]{Contini2021a,Montes2022}.
The ICL might form from several channels, but mostly arises from the gravitational 
interactions and merging between the galaxies \citep{2007MNRAS.377....2M}, which includes 
the tidal stripping of satellites, the disruption of dwarf galaxies
\citep[see e.g.,][]{Rudick2009,Contini2014,Contini2019}, and the pre-processing in groups \citep{Mihos2005,Rudick2006, Joo2023, Contini2024a}; for stellar population differences between the extended BCG haloes and the ICL, see the review by \citet{2022FrASS...972283A}).
As a consequence, a larger amount of ICL is expected in more evolved environments, that is, when galaxies have experienced many interactions.  

On the observational side, the detection and analysis of these structures are the most challenging tasks because stellar haloes and the ICL are diffuse and extremely faint. 
With their the long integration times and incrementally larger covered 
areas, the focused deep multi-band imaging and spectroscopic surveys in the past two decades have delivered data with a depth ($\mu_g \simeq 28-31$~mag/arcsec$^2$) and resolution that enabled the detection of galaxy outskirts and ICL. The extensive analyses 
of the light and colour distributions, kinematics, and stellar populations in different environments have been presented in many studies
\citep[see e.g.][and references therein]{Duc2015,Iodice2016,Mihos2017,Spavone2020,Danieli2020,Trujillo2021,Miller2021,Spavone2022,Gilhuly2022}. 
The comparison of the observations with theoretical predictions provides new ways to study the mass assembly in these different environments.  

Differently from simulated images, the ex-situ component cannot be unambiguously separated from the in-situ component based on deep observations alone. At the transition radius, both contribute equally to the light. In addition, at larger galactocentric distances, the ICL may also contribute to the stellar halo light, and the two may have indistinguishable surface brightness and colours.

To overcome this limitation, the widely adopted method for setting the size scales of the main
components that dominate the light distribution as a function of radius is the multi-component
fit of the azimuthally averaged surface brightness profiles 
\citep[see e.g.][and reference therein]{Trujillo2016,Spavone2017,Zhang2019,Kluge2020,Gonzalez2021,Montes2021}.
Using this approach, a lower limit to the amount of the accreted mass in a galaxy is derived 
by reproducing the light distribution out to the faintest levels. This value was derived
for a statistically significant sample of galaxies in groups and clusters and was found
to be consistent with the theoretical predictions, where a higher 
 accreted mass (i.e. the ex-situ component), which reaches 80\%-90\% of the total light,
is found in massive galaxies, that is, galaxies with a stellar mass of $\sim 10^{12}$~M$_\odot$ \citep[see][]{Spavone2020}. 

The measured radial velocities of the discrete tracers such as GCs and planetary nebulae (PNe) 
are a powerful tool for estimating the transition 
radius between the bound stellar haloes to the unbound envelope and, eventually, the ICL
\citep[e.g.][]{Coccato2013,Longobardi2013,Forbes2017,Spiniello2018,Hartke2018,Hartke2020}.
In particular, the number density distribution of the 
GCs and PNe at larger distances from the galaxy centre was found to follow the shape of the light profile of
the stellar envelope, which is shallower than the light profile in the central regions of the parent galaxy \citep[see e.g.,][]{Longobardi2015,Iodice2016}. 
In addition, the spatial projected distribution of the blue GCs correlates with the detected
ICL in clusters of galaxies 
\citep[][]{Durrell2014,Dabrusco2016,Iodice2017,Madrid2018,Cantiello2020}.

The most challenging task is the detection of the ICL in deep images. This requires
an ad hoc observing strategy to determine the best estimate of the background and the modelling of the light from the brightest cluster members.
To date, several methods have been efficiently used to detect the ICL in groups and clusters using ground- and space-based telescopes \citep[see][]{Montes2022, Montes2022a, Brough2024}.
Based on this, several observational properties of the ICL were derived, such as the 2D projected
distribution, colour, and metallicity, which are used to distinguish between the main 
formation channels of this component \citep[see][]{Iodice2017b,Mihos2017,DeMaio2018,Montes2021,Ragusa2021}.
\citet{Montes2019} showed that the ICL can be used as the luminous tracer of the DM. 

However, one of the most debated issues is the correlation between the
fraction of ICL  and the virial mass of the host environments and the scatter around it, which can provide useful information on the dynamical state of the host (e.g.
\citealt{Contini2024b}).
According to several theoretical works, the fraction of ICL ($f_{ICL}$) does not show
any trend. It ranges from 20\% to 40\% in the halo mass range M$_{vir} \simeq 10^{13}-10^{15}$~M$_\odot$ \citep[see][and references therein]{Rudick2011,Contini2014}. 
Conversely, increasing values of $f_{ICL}$ from 20\% up to 40\% have been predicted with increasing M$_{vir}$ \citep{Pillepich2018}, but the 
opposite trend was also predicted,  where a decreasing $f_{ICL}$ from $\sim$ 50\% 
to  $\sim$ 40\% with increasing M$_{vir}$ was found \citep{Cui2013}.
Based on the increased statistics in the ICL detection over a wide range of virial masses ($\sim$ $10^{12.5} \leq M_{vir} \leq 10^{15.5} M_{\odot}$), observational evidence emerges that $f_{ICL}$ and M$_{vir}$ of the host environment are not significantly correlated \citep[][]{Ragusa2023}.

We focus on a nearby galaxy cluster, Hydra I (see Sect.~\ref{sec:Hydra}), 
for which we map the light distribution down to the LSB regime in order
to detect and study the stellar haloes, including any remnant feature resulting from the gravitational interactions, and the ICL.
The main goal of this work is to address the assembly history of this cluster 
by studying the structure of the galaxy outskirts, constraining the accreted mass fraction
of the brightest cluster member (BCG), and estimating the ICL fraction.
In the following section, we provide a short review on the main properties of the Hydra I cluster.
In Sects.~\ref{sec:data} and~\ref{sec:phot} we present the data and the analysis we
performed, respectively. Our results are described in Sect.~\ref{sec:results} and discussed in Sect.~\ref{sec:discussion}.

\section{Hydra I  cluster of galaxies}\label{sec:Hydra}

Hydra~I is a rich and massive \citep[M$_{dyn} \sim 10^{14}$~M$_\odot$,][]{Girardi1998} cluster 
of galaxies located at $51\pm 4\;Mpc$ \citep[][]{Christlein2003}.
According to \citet[][]{Richter1982} and \citet{Richter1987}, Hydra~I is quite 
isolated in redshift space for a range of $40-50$~Mpc in front and behind the cluster 
along the line of sight. 
The X-ray emission detected out to $\sim 200$~kpc from the cluster core is centred on the brightest cluster galaxy (BCG)
NGC~3311, but is slightly displaced north-west \citep[NW,][]{Hayakawa2004,Hayakawa2006}.

The region of the cluster core was widely studied in the past ten years. 
It is dominated by the light of the BCG, NGC~3311, and the other early-type
bright member NGC~3309, which lies close in projection to NGC~3311 on the NW side.
Previous studies of the light distribution of these two galaxies suggested that they are both 
embedded in a diffuse and extended stellar envelope that is off-centred in the NE direction
\citep[][]{Arnaboldi2012}.
By combining the light distribution and stellar kinematics of the Hydra I cluster core, 
several works revealed 
i) stellar debris in the stellar envelope, which might result from the disruption of infalling dwarf galaxies,
ii) multiple components in the line-of-sight velocity dispersion of the PNe, suggesting
the existence of several different dynamical structures, and of iii) higher velocity
dispersion than the central galaxy regions, indicating that stars are gravitationally driven 
by the cluster potential \citep{Arnaboldi2012,Ventimiglia2011,Richtler2011, Hilker2018,Barbosa2021}.
All these features are signs of an active mass assembly around NGC~3311.
In particular, the velocity displacement measured between NGC~3311 and the barycentre 
suggests that the central part of NGC~3311 sloshes with respect to its outskirt. This is 
a clear sign of sub-cluster merging \citep{Barbosa2018}.

Using the Widefield ASKAP L-band Legacy All-sky Blind surveY (WALLABY) pilot survey with the Australian Square Kilometre Array Pathfinder (ASKAP), which covers the Hydra I cluster 
out to $\sim4$~Mpc ($\sim 2.5$~R$_{vir}$), \citet{Wang2021} found that nearly two-thirds 
of the galaxies within 1.25~R$_{vir}$ show signs of ram pressure stripping (RPS) in the early stages.
Two of these galaxies, NGC~3312 and NGC~3314A, are located SE of the core, 
in a foreground group of galaxies, and the RPS was also
studied in detail using MeerKat data by \citet{Hess2022}.

The Hydra~I cluster hosts a large population of dwarf galaxies. There are 317 candidates
in the magnitude range $-18.5 < M_r < -11.5$ mag. The first sample of
dwarf galaxies was provided by \citet{Misgeld2008} and was detected in a small area around 
the cluster core. Using a larger and deep-imaging mosaic, \citet{LaMarca2022a} found more than
200 new candidates, 32 of which are LSB galaxies, including ultra-diffuse galaxies (UDGs; \citealt{Iodice2020,LaMarca2022b}). Integral-field spectroscopic data
for all the LSB galaxies have recently been obtained with MUSE at the ESO-VLT
\citep[][]{Iodice2023}.

According to the 2D projected distribution of all cluster members, 
that is, bright and dwarf galaxies, three main overdensities were identified inside 0.4~R$_{vir}$
in the Hydra~I cluster. These are the core, a wider group toward the N, and a structure SE of the cluster core, while the western side of the cluster shows a lower density 
than the eastern side \citep{LaMarca2022a}. The study by \citet{LaMarca2022a} also revealed that the 
fraction of dwarf galaxies decreases toward the cluster centre. This is consistent with the tidal disruption process of the dwarfs in the strong potential well of the cluster core
\citep[see][and references therein]{Popesso2006}.
The presence of several sub-groups of galaxies in the Hydra I cluster further supports 
the scenario that this environment is still in an active assembly phase, where these structures merge into the cluster potential.


\section{Observations: Deep optical mosaics for the Hydra I cluster} \label{sec:data}

The deep $g-$ and $r-$band images of the Hydra I cluster presented in this work were acquired with the  VLT Survey Telescope \citep[VST,][]{Schipani2012} as part of the VST Early-Type
Galaxy Survey \citep[VEGAS,][]{iodice2021vegas}. VST is now a hosted telescope of the Italian 
National Institute for Astrophysics\footnote{See https://vst.inaf.it/home}, 
located at La Silla-Paranal European Southern Observatory (ESO) in Chile. The VST is equipped 
with the wide-field imager OmegaCAM \citep{kujiken2011}, 
which covers a field of view of $1\times1$~deg$^2$ and has a resolution of 0.21~arcsec/pixel.

Observations and data reduction have been described in \citet{Iodice2020c,Iodice2021} and \citet{LaMarca2022a,LaMarca2022b}.
We report the main information for this data set here.
The total integration times are $2.8$ hours in the $g$ band and $3.22$ hours in the $r$ band, respectively.
Images were acquired in dark time using the step-dither observing strategy, which guarantees an accurate estimate of 
the sky background \citep[see also][]{Iodice2016,Venhola2018}. The sky field was acquired west (W)
of the cluster. This field overlaps by $\sim0.3$ deg with the 1 deg field centred on the core of the cluster.
During the data acquisition, the bright (magnitude 7) foreground star NE of the cluster core 
was always placed in one of the two wide OmegaCAM gaps in order to reduce the scattered light. 

Pre-reduction, calibration, sky-subtraction, and stacking were provided by using the {VST-Tube}, 
which is one of the dedicated pipelines for processing OmegaCAM images \citep{Capaccioli2015}.
The final sky-subtracted reduced mosaic for the Hydra I cluster extends over 1 deg $\times$ 2 deg, 
corresponding to $\sim 0.9\times1.8$~Mpc, assuming a distance of 51 Mpc \citep[][]{Christlein2003}. 
Therefore, the VST mosaic covers $\sim 0.4$ virial radius around the core of the cluster.
The final stacked images reach surface brightness depths of $\mu_g=28.6\pm0.2$~mag/arcsec$^2$ and 
$\mu_r=28.1\pm0.2$~mag/arcsec$^2$ in the $g$ and $r$ bands, respectively.
The median FWHM of the point spread function is 0.85~arcsec in the $g$ band and 0.81~arcsec in the $r$ band. 

We focus on an area of $56.7\times 46.6$~arcmin$^2$ 
around the cluster core (Fig.~\ref{fig:mosaic}), which corresponds to $\sim 0.4R_{\rm vir}$.
Based on the \citet{Christlein2003} catalogue, this area includes 44 brightest ($m_R \leq 15$~mag) cluster members. They are listed in Table~\ref{tab:sample} and marked in Fig.~\ref{fig:mosaic}.

\section{Data analysis: Surface photometry}\label{sec:phot}

The main goal of this work is to study the outskirts of the brightest cluster members down to the 
LSB regime and to constrain the ICL fraction in this region of the cluster.
To do this, we studied the light distribution in the $g$ and 
$r$ bands for all galaxies in our sample (see Table~\ref{tab:sample}). We adopted the same methods and tools as previously developed within the VEGAS project for studies of the LSB regime and detection of ICL \citep[see][and references therein]{Iodice2016,Iodice2017,Spavone2018,Cattapan2019,Raj2019,Iodice2019}. 
A detailed description was provided by \citet{Ragusa2021}.
In short, the main steps of the surface photometry for the Hydra~I sky-subtracted mosaic are the following: 

\begin{enumerate}
    \item remove the contamination of the light from the foreground brightest stars in the field and background galaxies
    by modelling and subtracting it from the parent image;
    \item estimate the limiting radius of the photometry (R$_{lim}$) and the residual background fluctuations;
    \item fit the isophote  out to R$_{lim}$ for all the brightest galaxy members to obtain the azimuthally averaged surface brightness profiles and shape parameters;
    \item build the 2D model based on the isophote fit for each galaxy in our sample to be subtracted from the parent image in order to estimate the diffuse light in the residual image.

\end{enumerate}

The tools and methods adopted in each step are briefly described below.

\subsection{Contamination of the scattered light from stars}\label{sec:stars}

Close to the core of the Hydra~I cluster lie two bright stars: HD92036 (R.A.=10:37:13.72 DEC=-27:24:45.49, $m_B=6.51$~mag), 
located NE of NGC~3311, and HD91964 (R.A.=10:36:42.57 DEC=-27:39:22.86, $m_B=8.17$~mag), 
located south (S) of the core. During the observations, the centre of the brightest star HD92036 was always placed in one of the two wide
OmegaCAM gaps in order to reduce its scattered light. 
For both objects, we fitted the light distribution out the edge of the
field ($\sim 0.5$~deg; Fig.~\ref{fig:mosaic}), 
assuming circular isophotes.
The centre, position angle (P.A.), and ellipticity were fixed during the run. 
As a preliminary step, we identified all the background and foreground objects (stars and galaxies) that are brighter than the
$2\sigma$ background level \citep[see also][]{LaMarca2022b}, which were then masked and excluded from the isophote fit. 
In particular, the core of the group was masked with a circular aperture 
out to $\sim$ 7.7 arcmin ($\sim$ 13 R$_{eff}$) from the BCG NGC 3311 \citep[][]{Arnaboldi2012}.
Based on the isophote fit, the 2D models of the two stars were built up 
and then subtracted from the $g$ and $r$ parent images. 
The resulting image is shown in Fig.~\ref{fig:mosaic}.

\subsection{Estimate of the residual background fluctuations}\label{subsubsec:Rlim}

We overplotted on the star-subtracted residual image in each band the light distribution from the core of the cluster (identified 
as the centre of NGC~3311) out to the edge of the field ($\sim 0.5$~deg) in order to estimate the average value of the 
residual background level and the limiting radius R$_{lim}$ for the photometry in this field.
The residual background in the sky-subtracted images is the deviation from the background with 
respect to the average sky frame obtained from the empty fields observed close to the target. Therefore, 
since the reduced mosaics are sky-subtracted, the residual background level is close to zero in both bands.
As explained in \citet{Ragusa2021}, 
light is fitted in circular annuli (i.e. the ellipticity and PA 
are fixed to zero), with a constant step, where all the foreground and background sources, including the second BCG of Hydra I, NGC~3309, which is close to NGC~3311, were masked accurately.
R$_{lim}$ corresponds to the outermost semi-major axis derived 
in the isophote fitting, where the light from NGC~3311 blends into the average residual background level. The residual background fluctuations are almost constant for $R \geq R_{\rm lim}$ .
We found that for NGC~3311, R$_{lim}\sim 15$ arcmin in the $g$ and $r$ bands. 
For $R\geq R_{\rm lim}$, the residual background levels are $I_g = -0.056\pm 0.11$ ADU and 
$I_r = -0.6 \pm 0.2$ ADU in the $g$ and $r$ band, respectively. 
The residual background value and $R_{\rm lim}$ were derived for all galaxies in our sample in the $g$ and $r$ bands.

\subsection{Isophote fitting}\label{sec:isoph}

For all galaxies in our sample, we fitted the isophotes 
with all parameters left free (i.e. centre, ellipticity, and P.A.). The light distribution was mapped over elliptical annuli 
by applying a median sampling and k-sigma clipping algorithm. 
This iterative process was performed as follows. Firstly, we made the isophote fit for NGC~3311 and built a 2D model 
of the light distribution. 
This was subtracted from the parent image.
The isophote fit of the other bright cluster member in the core, NGC~3309, was performed on the residual image obtained before, 
and the 2D model for this galaxy was derived and subtracted. In turn, all the galaxies in our sample were studied with the same
approach, that is, the isophote fit was performed on the residual image obtained by subtracting the 2D model of the other galaxies 
studied in the sample.

The main outcome of the isophote fit is the azimuthally averaged surface brightness profile, which was corrected 
for the residual value of the background estimated for $R\geq R_{\rm lim}$. 
The total uncertainty on the surface brightness profile ($err_{\mu}$)\footnote{As detailed in \citet{Seigar2007,Capaccioli2015,Iodice2016}, the total uncertainty is derived as 
$err_{\mu}= \sqrt{(2.5/(adu \times \ln(10)))^2 \times ((err_{adu}+err_{sky})^2) +  err_{zp}^2}$, where
N is the number of pixels used in the isophote fit, adu is the analog digital unit and $err_{adu}=\sqrt{adu/N-1}$.}
took the uncertainties on the photometric calibration ($err_{zp}\sim0.3\%$) and the RMS in the background fluctuations into account
($err_{sky} \sim10\%$ and $err_{sky} \sim20\%$ in the $g$ and $r$ bands, respectively).
In addition, we derived the $g-r$ colour profiles, the total integrated magnitudes, and the $g-r$ integrated colours. 
The surface brightness and colour profiles for NGC~3311 are shown in Fig.~\ref{fig:prof_3311}. 
Those obtained for all the other galaxies in our sample are shown in Appendix~\ref{sec:sb_prof}. 
The total integrated magnitudes and the $g-r$ integrated colours are listed in Table~\ref{tab:sample}.

\begin{table*}
\setlength{\tabcolsep}{1.8pt}
\begin{center}
\small 
\caption{Photometric properties and redshift of the brightest cluster members inside $0.4R_{\rm vir}$ of 
the Hydra~I cluster.} 
\label{tab:sample}
\vspace{1pt}
\begin{tabular}{lcccccccccccc}
\hline\hline
Object  &  R.A.  &  Dec.  &  V  &  $R_{\rm proj}/R_{\rm vir}$  &  $m_{g}$ &  (g-r)  &  Morph.  &  Names\\
  & [J2000]   &[J2000]   &  [km/s]  &   &  [mag]  &  [mag] &   &  \\
 (1)   &  (2)  & (3) &  (4)  &  (5)  &  (6) &  (7)  & (8)  &  (9) \\
\hline \vspace{-7pt}\\
NGC~3311&10 36 42.70&-27 31 42.00&3857&0&10.3$\pm$0.11&0.6$\pm$0.3&cD2&HCC001\\
NGC~3316&10 37 37.26&-27 35 38.50&3922&0.012&12.81$\pm$0.04&0.87$\pm$0.09&SB(rs)0&HCC004\\
NGC~3309&10 36 35.69&-27 31  5.30&4071&0.016&12.08$\pm$0.05&0.83$\pm$0.11&E3&HCC002\\
SGC~1034.3-2718&10 36 41.15&-27 33 39.10&4735&0.018&14.52$\pm$0.01&0.80$\pm$0.04&S0&HCC007\\
PGC~031407&10 36  04.20&-27 30 31.30&2294&0.032&14.59$\pm$0.01&0.79$\pm$0.02&cD pec&\\
PGC~031483&10 36 44.89&-27 28  9.80&2735&0.033&14.30$\pm$0.01&0.78$\pm$0.02&E1&HCC006\\
LEDA~087333&10 36 35.45&-27 28 11.90&3222&0.036&15.49$\pm$0.01&0.78$\pm$0.02&E&HCC013\\
PGC~031450&10 36 29.04&-27 29  2.40&4774&0.037&14.54$\pm$0.04&0.84$\pm$0.09&SB&HCC009\\
NGC~3312&10 37  02.53&-27 33 53.60&2761&0.045&11.92$\pm$0.01&0.65$\pm$0.03&SA(s)b pec&\\ 
PGC~031444&10 36 24.86&-27 34 53.90&2788&0.047&14.76$\pm$0.01&0.74$\pm$0.03&E2&HCC011\\
NGC~3307&10 36 17.12&-27 31 46.50&3773&0.053&14.47$\pm$0.02&0.86$\pm$0.04&SB&\\
NGC~3308&10 36 22.31&-27 26 17.50&3537&0.066&12.27$\pm$0.02&0.93$\pm$0.04&SAB0&HCC003\\
PGC~031418&10 36 10.94&-27 27 14.50&4937&0.077&14.71$\pm$0.01&0.74$\pm$0.02&S0&\\
ESO~501-G047&10 37 17.01&-27 28  7.60&4821&0.078&12.26$\pm$0.01&0.96$\pm$0.23&SB0&\\
ESO~501-G049&10 37 19.95&-27 33 33.70&4020&0.079&14.33$\pm$0.02&0.27$\pm$0.07&S0&HCC008\\
PGC~031464&10 36 34.95&-27 28 44.30&4691&0.080&15.04$\pm$0.02&0.79$\pm$0.05&S0&HCC012\\
PGC~031515&10 37  04.89&-27 23 59.30&2690&0.085&14.59$\pm$0.01&0.39$\pm$0.01&S0&\\
ABELL~1060:SMC-S135&10 37  09.62&-27 39 27.90&4126&0.090&14.00$\pm$0.04&0.56$\pm$0.09&E3&\\
PGC~031402&10 35 57.90&-27 33 43.90&3571&0.094&14.89$\pm$0.02&0.77$\pm$0.05&S0&\\
PGC~031441&10 36 23.07&-27 21 14.80&3005&0.105&14.47$\pm$0.03&0.76$\pm$0.08&S0&HCC010\\
NGC~3314&10 37 12.76&-27 41  1.10&2795&0.105&12.93$\pm$0.01&0.50$\pm$0.02&S&\\
LEDA~101367&10 36 37.50&-27 43  0.90&3828&0.106&15.33$\pm$0.02&0.52$\pm$0.05&S0&\\
PGC~031422&10 36 12.96&-27 41 18.60&4132&0.108&14.75$\pm$0.01&0.75$\pm$0.03&S0&\\
PGC~031432&10 36 18.92&-27 43 16.70&3559&0.118&15.11$\pm$0.02&0.81$\pm$0.04&SB&\\
PGC~031447&10 36 27.64&-27 19  8.50&3376&0.119&13.99$\pm$0.02&0.78$\pm$0.04&S0&HCC005\\
ABELL~1060:[R89] 196&10 35 38.27&-27 31 34.30&3026&0.132&15.31$\pm$0.04&0.82$\pm$0.11&S&\\
ESO~501-G052&10 37 36.80&-27 23 14.00&3633&0.136&14.19$\pm$0.11&1.04$\pm$0.23&S0&\\
ESO~501-G027&10 35 57.86&-27 19  7.90&3158&0.149&14.55$\pm$0.04&0.72$\pm$0.10&E6&\\
LEDA~087328&10 37 19.42&-27 16 23.70&4463&0.160&14.98$\pm$0.01&0.64$\pm$0.03&S0&\\
ESO~501-G026&10 35 24.69&-27 28 55.00&2954&0.163&13.99$\pm$0.12&0.40$\pm$0.32&S0&\\
PGC~031371&10 35 30.81&-27 22 49.30&4430&0.169&14.01$\pm$0.02&0.68$\pm$0.04&S0a&\\
ESO~501-G021&10 35 20.48&-27 21 42.90&4539&0.192&13.68$\pm$0.01&0.74$\pm$0.03&S0&\\ 
NGC~3315&10 37 19.17&-27 11 30.50&3753&0.201&13.17$\pm$0.01&0.77$\pm$0.02&S0&\\
LEDA~141477&10 36 35.60&-27  08 56.20&4686&0.211&15.15$\pm$0.03&0.80$\pm$0.05&E/S0&\\
NGC~3305&10 36 12.04&-27  09 43.20&4002&0.212&13.02$\pm$0.01&0.81$\pm$0.01&E0&\\
ESO~501-G065&10 38 33.32&-27 44 12.40&4412&0.255&13.27$\pm$0.01&0.28$\pm$0.02&SB(s)d&\\
ESO~501-G041&10 36 53.04&-27  03 10.60&3643&0.264&14.55$\pm$0.02&0.62$\pm$0.06&SB&\\
ESO~501-G059&10 37 49.40&-27  07 15.20&2434&0.264&13.11$\pm$0.05&0.71$\pm$0.12&Sc&HCG048B\\
NGC~3285B&10 34 36.75&-27 39  9.30&3150&0.268&13.28$\pm$0.02&0.50$\pm$0.04&SAB&\\
ABELL~1060:[R89] 185&10 35 12.11&-27 10 10.00&4292&0.272&15.00$\pm$0.03&0.66$\pm$0.04&S0&\\
IC~2597&10 37 47.30&-27  04 52.00&2973&0.281&11.75$\pm$0.02&0.84$\pm$0.05&cD4&HCG048A\\
HCG~048C&10 37 40.51&-27  03 28.20&3124&0.287&12.78$\pm$0.45&1.22$\pm$0.82&S0a&\\
ESO~501-G020&10 34 47.70&-27 12 51.50&4369&0.294&13.31$\pm$0.20&1.04$\pm$0.38&SB0&\\
HCG~048D&10 37 41.52&-27  02 39.40&4326&0.294&14.98$\pm$0.02&0.78$\pm$0.05&E1&\\

\hline
\end{tabular}
\tablefoot{
(1)--(4) Galaxy name, RA, Dec, and heliocentric velocity were taken from \citet{Christlein2003}. 
(5) Ratio of the projected distance from the cluster centre and virial radius of the cluster ($R_{\rm vir}=1.6$ Mpc).
(6)--(7) $g$-band total magnitude, $(g-r)$ colour, corrected for Galactic extinction \citep{Schlegel98}. (8) Morphological type (from the NED). (9) Alternative galaxy name.}
\end{center}
\end{table*}

\begin{figure*}
    \centering
    \includegraphics[width=18cm]{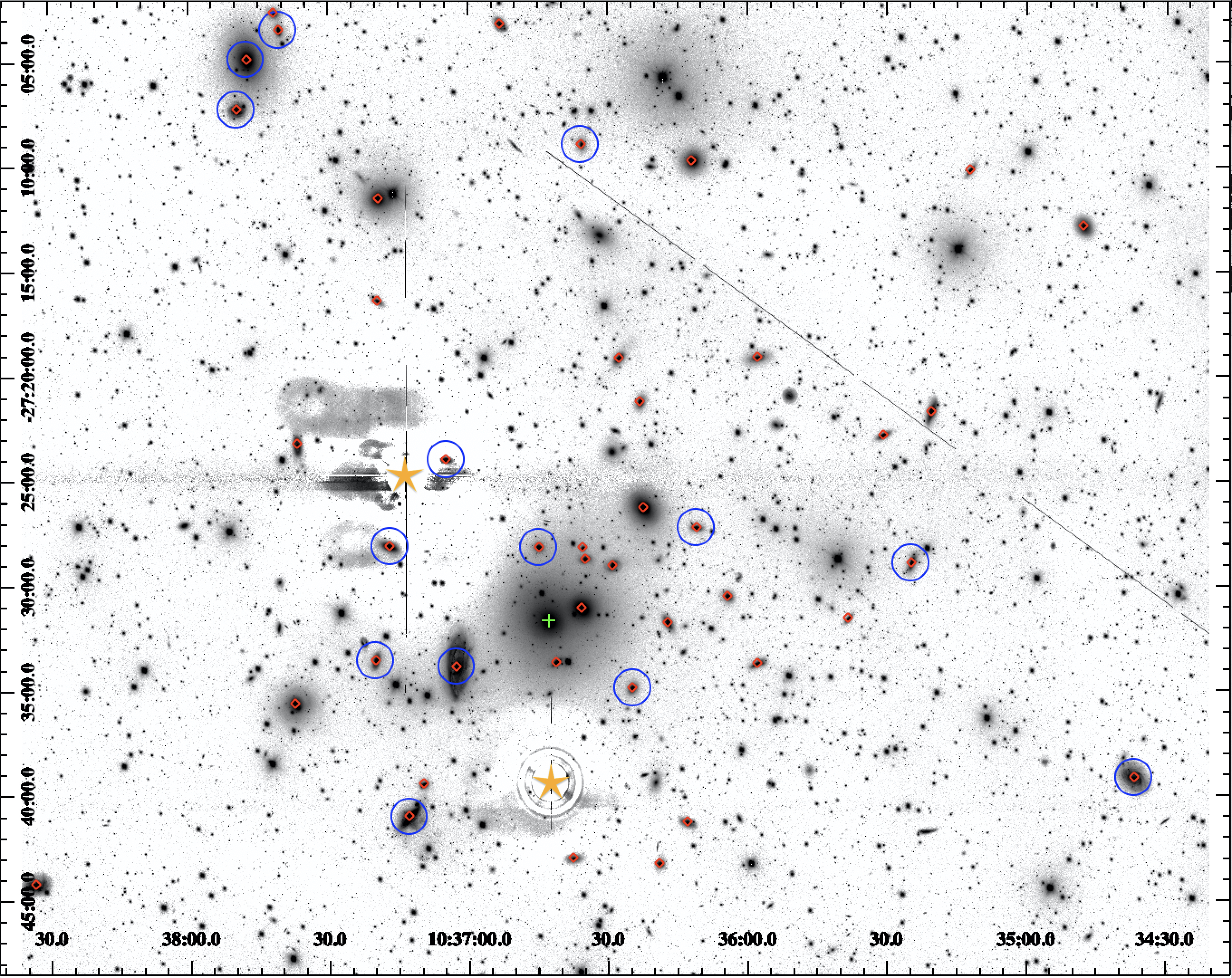}
    \caption{Extracted region of the g-band VST mosaic ($0.945 \times 0.776 \deg$ $\sim0.84 \times 0.69$~Mpc) of the Hydra I 
    cluster, which corresponds to $\sim 0.4 R_{\rm vir}$. 
    The brightest galaxy members ($m_R \leq 15$~mag) from the \citet{Christlein2003} catalogue are marked as red diamonds. The BCG, NGC~3311, is marked with the green cross. The large blue circles indicate the galaxies whose heliocentric velocity is comparable to or larger than the escape velocity of the cluster in phase-space (see Sect.~\ref{sec:discussion}). 
    The orange symbols mark the position of the two brightest foreground stars in the field, which were subtracted from the original mosaic (see Sect.~\ref{sec:isoph}). As expected, some artefacts are still present close to the subtracted stars, which are caused by non-symmetric scattered light and the haloes of those two brightest stars.
    N is up, and east (E) is on the left.}
    \label{fig:mosaic}
\end{figure*}

 \begin{figure*}
    \centering
        \includegraphics[width=9.2cm]{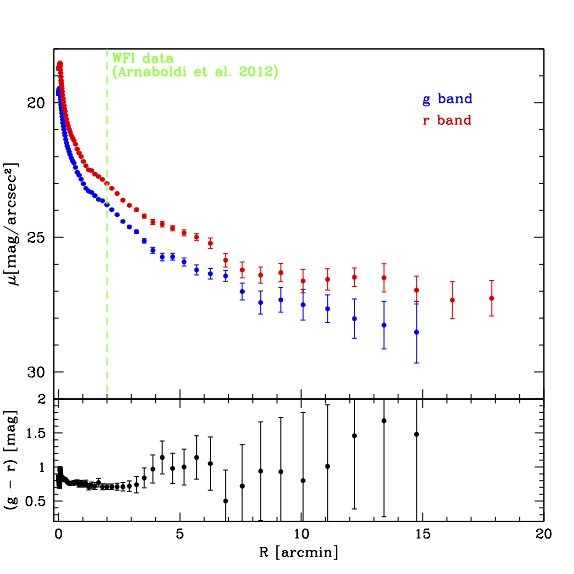}
    \includegraphics[width=8.7cm]{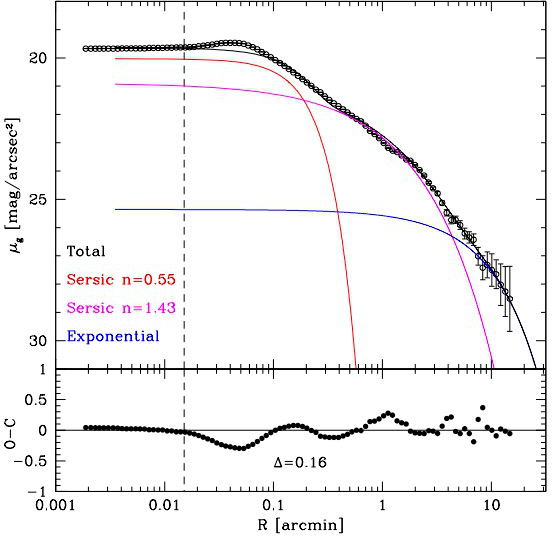}
    \caption{Azimuthally averaged surface brightness of NGC~3311. Left panel: Surface brightness profiles (top panel) in the $g$ (blue points) and $r$ (red points) bands, and $g-r$ colour profile (bottom panel). The dashed green line marks the limit of the surface brightness profiles obtained for NGC~3311 by \citet{Arnaboldi2012}. 
    Right panel: $g$-band surface brightness profile in logarithmic radius (top panel), with the best 1D multi-component fit (see text for details). In the lower panel, we show the $\Delta$rms residual of the data minus the model (see text for details).}
    \label{fig:prof_3311}
\end{figure*}


\subsection{Globular cluster number density}\label{sec:GCs}
For the Hydra I cluster, we collected a sample of $\sim5300$ GC candidates obtained from deep U-, V- and I-band 
photometry observed with VLT/FORS \citep{Mieske2005, Hilker2015} extending out 
to $\sim20$~arcmin N and E of NGC~3311.

\begin{figure*}
    \sidecaption
    \includegraphics[width=12cm]{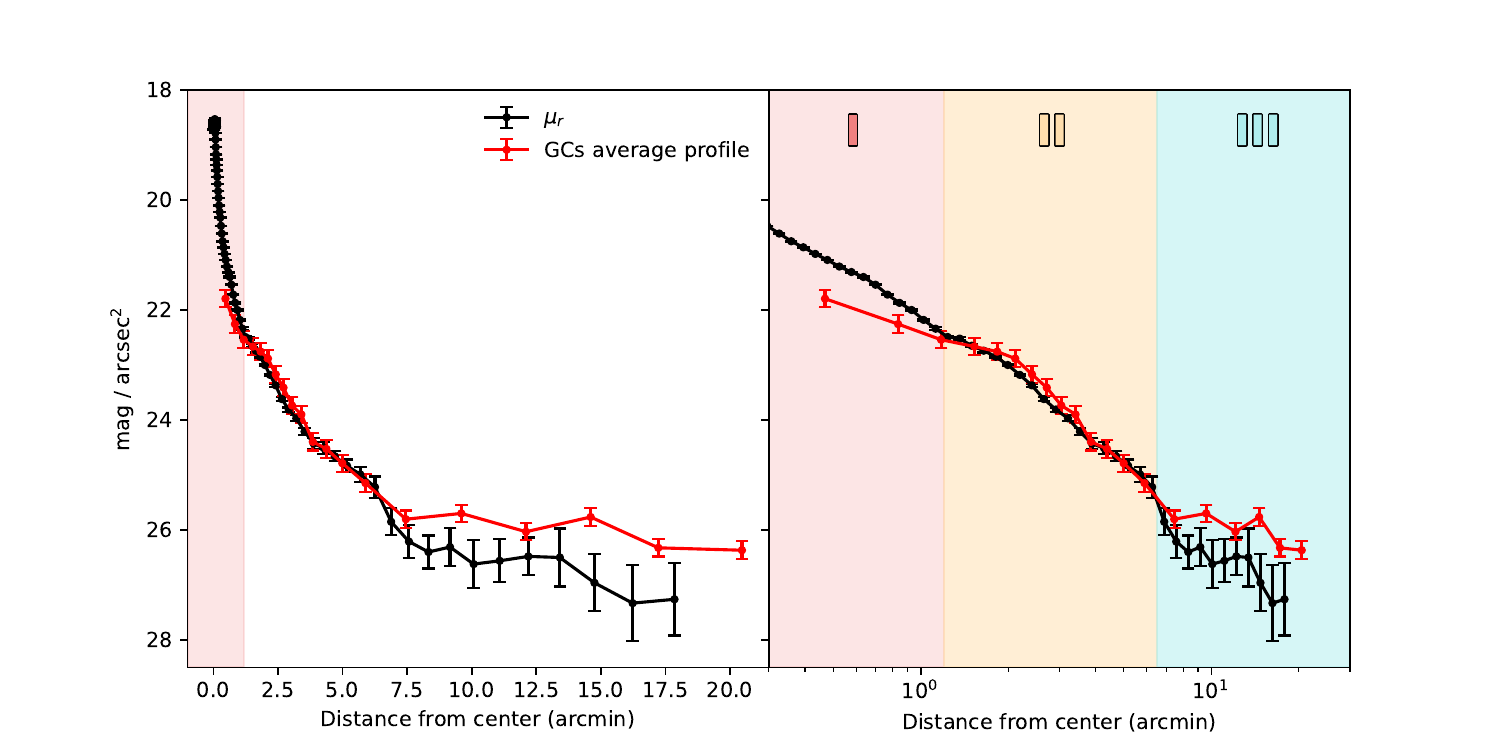}
   \caption{Surface brightness profile of NGC3311 in the r band (black line) and the radially averaged globular cluster number density profile (red line), arbitrarily shifted to match the surface brightness. The left panel shows the cluster-centric distance on a linear scale, and the right panel uses a log scale. The shaded areas indicate the three regimes observed in the GC number density profile: I) where incompleteness effects are significant, II) where the GCs follow the galaxy light, and III) where the GCs trace the ICL. Region I) is also shaded in the left panel for reference. At distances larger than $\sim 6$~arcmin from the centre, the shallower profiles indicate the contribution of the ICL baryons, i.e. diffuse light plus intra-cluster GCs.}
   \label{fig:GCprof}
\end{figure*}

The GCs were selected based on the morphology (point sources) and their location in the V - (V-I) colour-magnitude diagram, guided by radial-velocity-confirmed member GCs \citep{Misgeld2011}. The details of the data reduction and sample selection will be described in a future paper (Lohmann et al. in prep.).
We used the spatial distribution of GCs to obtain their average radial number density profile. 
The profile was constructed considering radial annuli centred on NGC~3311 such that each bin 
contained a fixed number of GCs. 
We masked out the GCs associated to NGC~3309 using a 
circular mask around the galaxy, as well as those forming an overdensity to the east (E), 
which are associated with the lenticular galaxy NGC~3316. 
The mask radius was arbitrarily chosen to be $0.9$ arcmin, which corresponds to $\sim$2.3 $R_e$ for NGC 3309 and  $\sim$3.5 $R_e$ for NGC 3316. 
By adopting 20 bins for our profile, we obtained $\sim250$ GCs in each bin, and the uncertainty 
on the number density was taken to be Poissonian. After correcting for background contaminants, 
the surface density of the globular clusters in units of arcsec$^{-2}$ was then shifted 
arbitrarily to match the surface brightness profile of NGC~3311. The results are shown in Fig.~\ref{fig:GCprof}.

The same sample selection steps included the analysis of a background region located $\sim$1.5 deg E
of Hydra~I to select GC-like objects that contaminate our sample. 
We calculated the number density of these objects and subtracted it from the GC number density 
profile. The GC sample was also corrected for incompleteness by examining the GC luminosity 
function in the V band, to which we fitted a Gaussian distribution and determined an absolute peak magnitude of $M_{V} = -7.4$~mag and $\sigma= 1.1$~mag. We then calculated a completeness factor by which the GC counts were to be multiplied, which was $\sim2.5$\footnote{The completeness factor was calculated using the Gaussian fit to the luminosity function of our GCs. We first integrate the function from $-\infty$ to $+\infty$ to get the total number of GCs that we expect from the LF. We then integrate the function from $-\infty$ to our limiting absolute magnitude (around -7.7) to get the expected number of observed GCs. We then divide the two numbers and get the factor of 2.5.}.
At small radii ($\leq$ 1.2 arcmin), the detection algorithm failed to identify sources due 
to the high background light from NGC~3311, leading to an incompleteness in the central regions.

\section{Results: Galaxy outskirts, diffuse light, and low surface brightness features}\label{sec:results}

\subsection{Cluster core} \label{sec:core}

\begin{figure*}
    \centering
    \includegraphics[width=18cm]{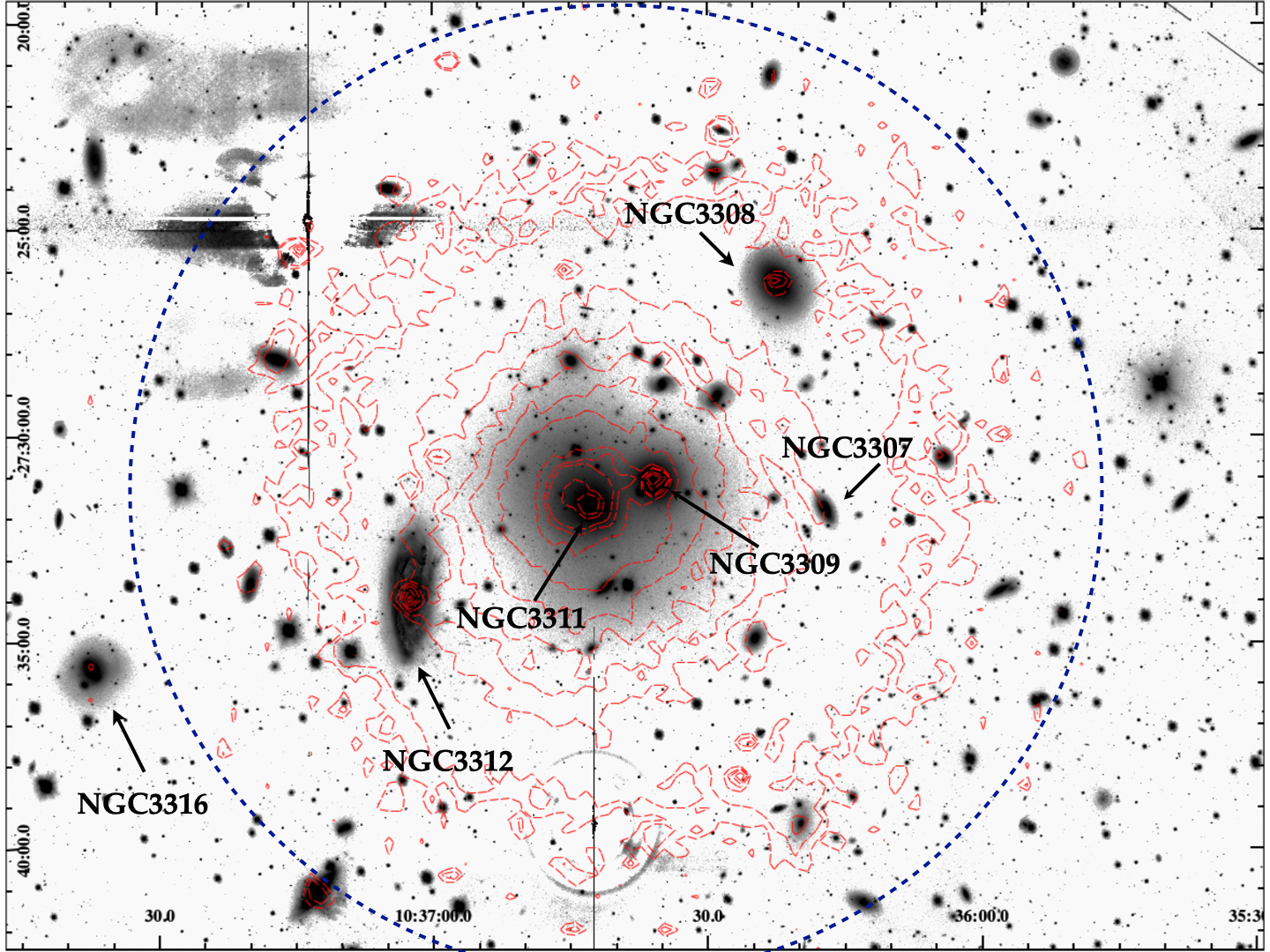}
    \caption{Enlarged region of the Hydra~I cluster centred on the core. N is up, and E is left. NGC~3311 and NGC~3309 dominate the cluster centre, while NGC 3312 is visible in the SE. This is member of the SE group. Another two bright galaxies in the core, NGC~3308 and NGC3307, are also marked in the image. The dashed red contours indicate the X-ray emission from XMM \citep[][]{Hayakawa2004,Hayakawa2006}. The dashed blue circle marks the outermost radius of 14~arcmin$\sim220$~kpc, where the surface brightness profiles are mapped, down to $\mu_g\sim28.5$~mag/arcsec$^2$, in the $g$ band.
    The image is $30.5\times23.0$~arcmin$^2$ wide.}
    \label{fig:core}
\end{figure*}

The core of the cluster is dominated by the extended diffuse envelope around the two brightest cluster galaxies, NGC~3311 and NGC~3309 (Fig.~\ref{fig:core}).
The light distribution was mapped out to 15 arcmin from the centre of NGC~3311, which corresponds to $\sim 223$~kpc ($\sim 0.14$~R$_{vir}$), and down to $\mu_g\sim28.5$~mag/arcsec$^2$ in the $g$ band (Fig.~\ref{fig:prof_3311}, left panel). This distance coincides with the extension of the detected X-ray emission, centred on the cluster core (Fig.~\ref{fig:core}). 
Our new surface brightness profiles are about six times more extended than those obtained in the optical $V$ band by \citet[][]{Arnaboldi2012}. 

\begin{figure*}
    \centering
        \includegraphics[width=1.\textwidth]{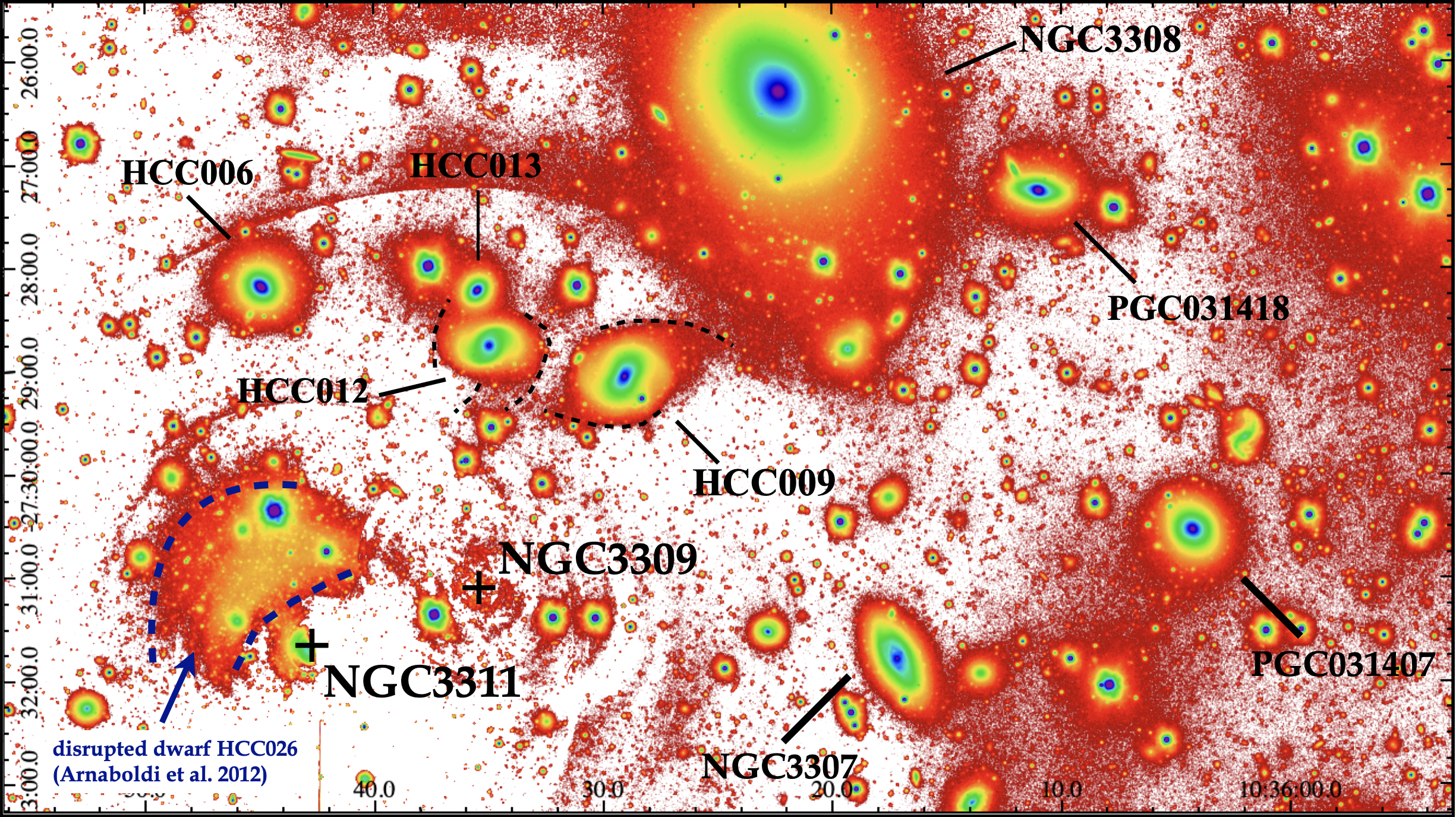}
    \caption{Enlarged region of the Hydra~I cluster on the NW side of the core. This is the residual image obtained by subtracting the 2D model of the light distribution of NGC~3311 and NGC~3309 from the isophote fit (see text for details). 
    The image is $14.06\times7.81$~arcmin$^2$ wide.
    The brightest galaxies in the field are marked. In addition, the dashed black lines indicate the distorted isophotes in the outskirts of HCC092 and HCC012, which might suggest possible ongoing interaction. 
    The dashed blue lines mark the stellar streams of the disrupted dwarf HCC026 identified by \citet[][]{Arnaboldi2012}.
    The several ripples and arcs are artefacts from the residuals of the overlapping outskirts of NGC~3311 and NGC~3309 with the numerous nearby galaxies, in particular, the bright member NGC~3308.}
    \label{fig:core_residual}
\end{figure*}

In Fig.~\ref{fig:core_residual} we show a portion of the residual image obtained by subtracting 
the 2D model of the light distribution of NGC~3311 and NGC~3309, derived from the isophote fit (see Sect.~\ref{sec:isoph}).
This region of the cluster is located NW of the core.
Several bright galaxies in this area show distorted isophotes in the outskirts. In detail, 
HCC009, which is a barred S0 galaxy, has a prominent S-shape in the west-east (WE) direction.
A similar morphology is observed for HCC012, which lies close in projection to HCC009 on the W side, where
LSB tails are detected N and S.
The giant galaxy NGC 3308, located north of the cluster core, which has a similar total magnitude as
NGC 3309 (see Table~\ref{tab:sample}), has an extended stellar envelope that is elongated toward the SW.
We also confirm the tidal debris associated with the disrupted dwarf HCC026 inside the stellar envelope of NGC~3311, which was discovered by \citet[][]{Arnaboldi2012}.

\subsection{Extended envelope around NGC~3311}\label{subsubsec:fit1d}

Previous studies of imaging data for NGC~3311 revealed that this galaxy is composed of a central and bright component that is well fitted by 
a Sersic law, and an extended stellar envelope \citep[][]{Arnaboldi2012}. 
As stated in the previous section, the surface brightness profiles obtained from the VEGAS data 
are about six times more extended than previous profiles. 
It shows a change in the slope at $\sim 4$~arcmin, where a shallower decline is observed (Fig.~\ref{fig:prof_3311}, left panel), suggesting an additional outer component.
Therefore, we performed a new multi-component 1D fit of the surface brightness profile for NGC 3311 in the $g$ band. We adopted the same approach as proposed in many studies that is also well tested for VST data to fit the surface 
brightness profiles of BGCs in the cluster of galaxies
\citep[see e.g.][and references therein]{Spavone2020}. In detail, 
the best fit was obtained with three components: two Sersic laws, which reproduce the 
brightest and inner regions of the profile, and an outer exponential profile that maps 
the galaxy outskirts. The results are shown in the right panel of Fig.~\ref{fig:prof_3311}.
The best-fit parameters are listed in Table~\ref{tab:1Dfit_3311}. 

The transition radius between the inner two Sersic components is $R_{\rm tr}\geq 0.2$~arcmin 
($\sim 3$~kpc). The outer exponential component starts to dominate at  $R_{\rm tr}\geq 4.3$~arcmin ($\sim$ 60 kpc).
It has a scale length of $\sim 74$~kpc and a total luminosity of
$\sim 1.6 \times 10^{11}$~L$_{\odot}$. This corresponds to $\sim 70$\% of the total luminosity 
integrated over the whole profile, which is $\sim 2.3 \times 10^{11}$~L$_{\odot}$.
This component traces the stellar envelope around the core of 
the cluster. We assume that it also includes the ICL in this region because
based on the photometry alone, we cannot separate 
the bound stars in the galaxy stellar halo from those that are gravitationally unbound in the ICL.
Based on the total luminosity of the two outer components, that is, the Sersic plus 
the exponential components, we estimated the accreted mass fraction
around NGC~3311, which is $f_{ac} \sim 94\%$. 
As argued by \citet{Remus2022}, because cosmological simulations only consider the stars that were already formed at the time of infall as accreted, the accreted fractions derived following the prescriptions of these simulations should be considered as lower limits. 
As explained in \citet{Spavone2017}, $f_{ac}$ includes the total luminosity of the
second Sersic component plus the outer exponential, which corresponds to the bound and
unbound ex situ components, respectively \citep[see also][]{Cooper2015}.

In correspondence with the  outer edge of the stellar envelope, that is, $R_{\rm lim}\sim15$~arcmin $\sim223$~kpc, we found that the number density of the dwarf galaxy population 
\citep[][]{LaMarca2022a} as a function of the cluster-centric distance decreases by about 10\% with
respect to the values at larger radii (Fig.~\ref{fig:Dwarf_ND}). 
A similar result was found for the Fornax cluster \citep{Venhola2018}.

\begin{table}[]
\setlength{\tabcolsep}{1.2pt}
\small
\begin{center}
    \caption{Best-fit parameters of the multi-component 1D fit of the surface brightness profiles of NGC~3311 in the $g$ band.}
    \begin{tabular}{lccccc}
    \hline
         Law & $\mu_e$ & $r_e$ & $n$ & $\mu_0$ & $r_h$ \\
                   & [mag/arcsec$^2$] & [arcsec] & & [mag/arcsec$^2$] & [arcsec]\\
          (1) & (2) & (3) & (4) & (5) & (6)\\         
         \hline\hline
         Sersic & 20.87$\pm$0.02 & 8.35$\pm$0.03 & 0.55$\pm$0.04 & - & -\\
         Sersic & 23.64$\pm$0.04 & 99$\pm$2 & 1.43$\pm$0.01 & - & -\\
         Exp & - & - & - & 25.36$\pm$0.01 & 298$\pm$6\\
         \hline
    \end{tabular}
    \tablefoot{Column 1 reports the empirical law adopted for the multi-component 1D fit.
    Columns 2 to 4 list the effective surface brightness, effective radius, and Sersic's index of the Sersic law for each of the two inner components. Columns (5) and (6) list the central surface brightness and scale length of the exponential component we adopted to fit the galaxy envelope.}
    \label{tab:1Dfit_3311}
    \end{center}
\end{table}


\subsection{Diffuse light versus distribution of the globular clusters}\label{sec:GCs_ICL}

Figure~\ref{fig:GCprof} shows the radial average number density profile of GCs around NGC~3311
(red line) along with the $r-$band surface brightness profile (black line). 
The right panel clearly reveals three different regimes of the number density of GCs, 
as indicated by the shaded regions.
Region I corresponds to the small-radius regime where the incompleteness effects in the GC counts 
are substantial. As discussed in Sect.~\ref{sec:GCs}, this is caused by the source detection 
algorithm, which underestimates the GC numbers. This is mostly caused by the GC number density profile, which traces the galaxy light at distances below $\sim$1.2 arcmin ($\sim17.8$ kpc) only poorly.
Region II spans distances between $1.2 \leq d \leq 6.5$~arcmin ($17.8 \leq d \leq 96.4$ kpc), 
where the number density of GCs closely follows the light from NGC~3311.
The stellar envelope also starts to dominate in this region \citep[see][and Sect.~\ref{subsubsec:fit1d}]{Barbosa2018}.

Region III ($d \geq 6.5$ arcmin) shows a striking difference between the light distribution and GC profile. At $R\simeq 6$~arcmin, the GC profile shows a shallower decrease than the surface brightness profile. At this radius, the extended exponential surface brightness profile starts to dominate (Fig.~\ref{fig:GCprof}). 
Therefore, region III not only marks a change in the surface brightness behaviour, but also 
in the mixture of stars and GC systems. 
The left panel of Fig.~\ref{fig:GCprof} shows that for $R\geq 6$~ arcmin, there are more GCs by factor of $\sim2.5$ than what is expected from the stars that contribute to the surface brightness profile. 
Therefore, at large distances, there are 2.5 times more GCs per unit galaxy light 
than in the inner region. This is a strong indication of the transition 
between the galaxy stellar halo and the ICL-dominated regime.
The connection between the flattening in the GC number density and this change in regime was also reported in previous works. \citet{Durrell2014} studied the GC population in the Virgo galaxies
M87 and M49 and found that their GC number density also flattens at large distances due to the 
shallow profile of blue GCs, which are connected to the ICL of Virgo. 
Moreover, studying planetary nebulae (PNe) in M87, \citet{Longobardi2015} found that 
intracluster PNe have a shallower profile than those belonging to the M87 halo. 
Similar results were obtained for the blue GCs around the central Fornax cluster galaxy NGC~1399
\citep[][]{Schuberth2010,Cantiello2018}.

\begin{figure}
    \includegraphics[width=9cm]{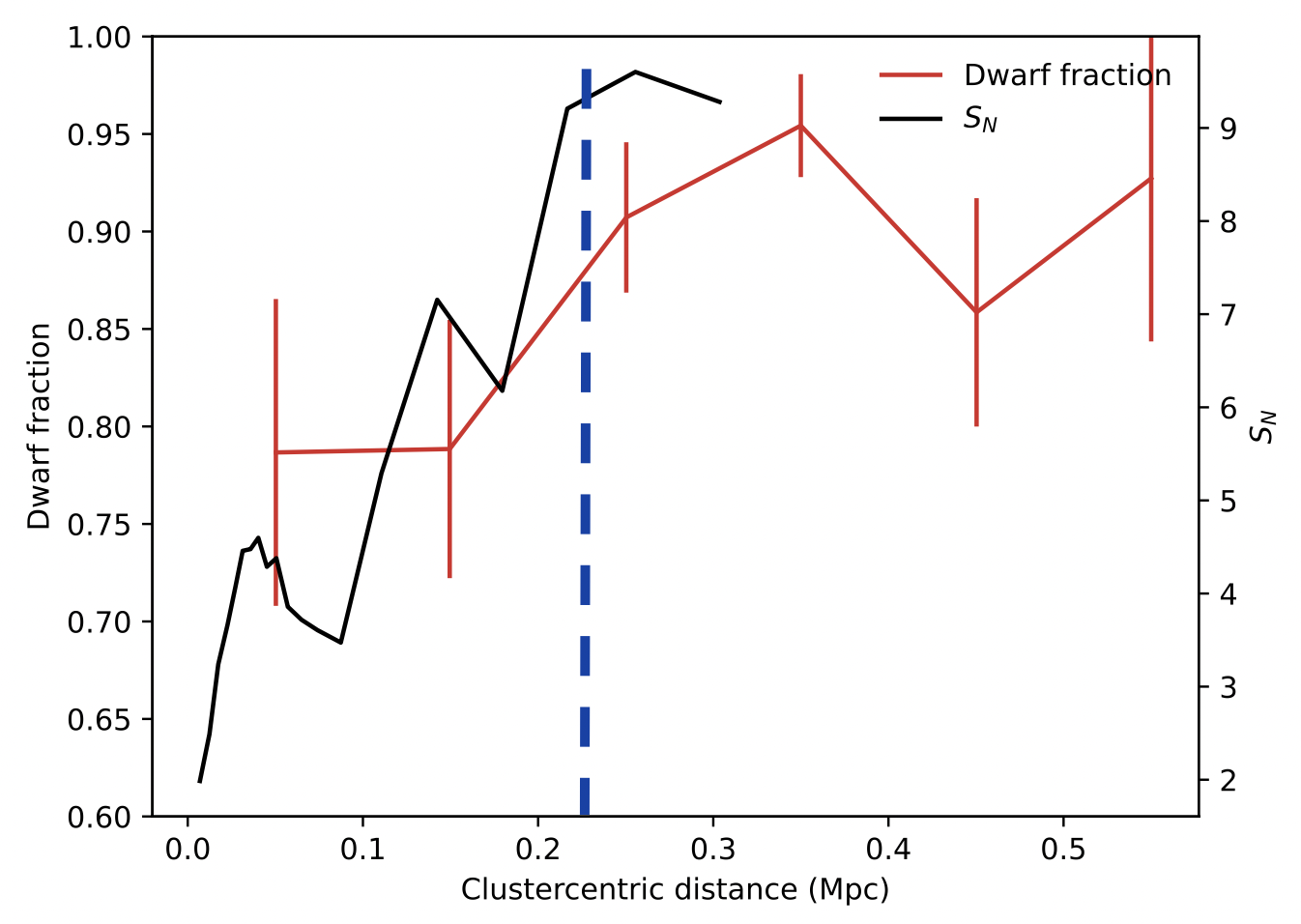}
    \caption{Number density of the compact sources as function of the cluster-centric distance.
    The number density (left axis) of the dwarf galaxy population (red line) is taken from \citet[][]{LaMarca2022a}.  The GC-specific frequency (black line, right axis) was derived in this work and is described in Sect.~\ref{sec:GCs_ICL}.
    The vertical dashed blue line marks the outer edge of the stellar envelope around the core, i.e. $R_{\rm lim}\sim15$~arcmin $\sim223$~kpc (see Fig.~\ref{fig:prof_3311}).}
    \label{fig:Dwarf_ND}
\end{figure}

Finally, we derived the specific frequency $S_N$ profile using the GC number density 
and the surface brightness profiles by integrating them in each annulus.
This is not the classical definition of $S_N$, which typically considers the total galaxy luminosity in the calculation, while we only considered the light inside each respective annulus here.
The $S_N$ profile as a function of cluster-centric distance is plotted in Fig.~\ref{fig:Dwarf_ND}, 
which shows that at larger distances from NGC~3311, an excess of GCs is observed, 
with a specific frequency that is about four times higher than in the central regions. 
A similar trend is observed for the number density of the dwarf galaxies, as discussed in Sect.~\ref{sec:core}.

\subsection{Disrupting dwarf in the outskirt of NGC~3316}\label{sec:NGC3316}

\begin{figure*}
    \centering
    \includegraphics[width=1.\textwidth]{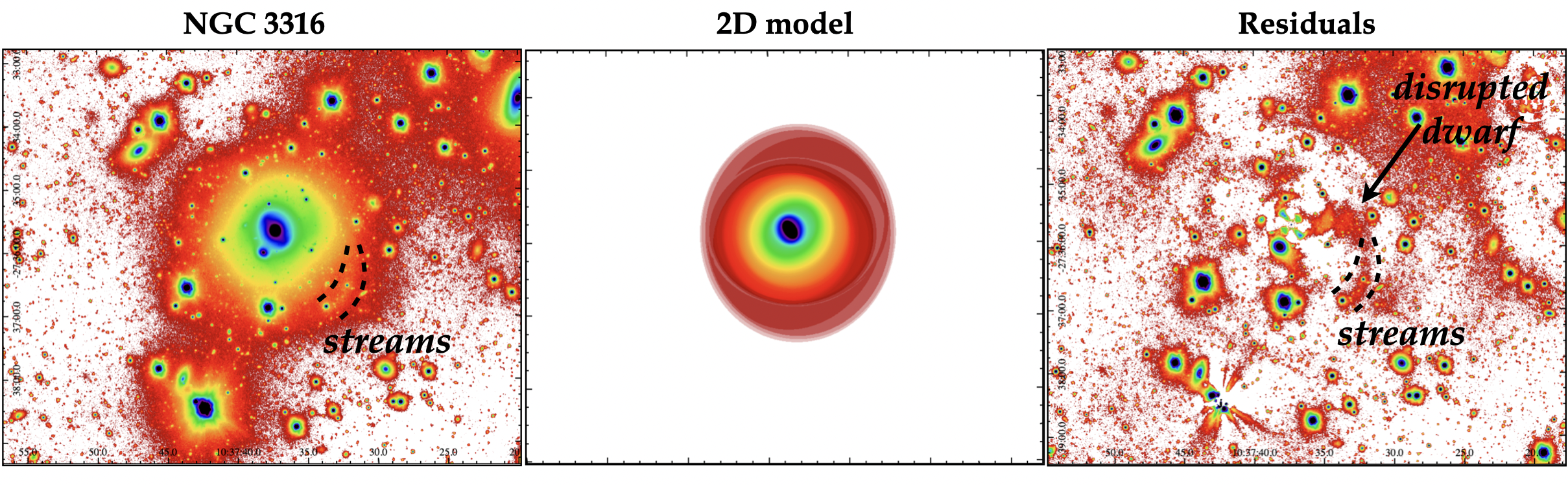}
    \caption{Outskirts of the early-type galaxy NGC~3316, located E of the core.
    The g-band region centred on NGC~3316 is shown in the left panel. We mark the prominent stellar stream on the SW side of the galaxy. The 2D model obtained from the isophote fit is shown in the middle panel. The right panel shows the residual image, where the 2D model (middle panel) is subtracted from the parent image (left panel). The stellar stream seems to be connected with a bright knot on top of it. Both features might be the remnant of dwarf galaxy, disrupted in the potential well of NGC~3316. }
    \label{fig:NGC3316}
\end{figure*}

On the SE side of the cluster core lies the fifth brightest member, NGC~3316 (Fig.~\ref{fig:core}),
which has a systemic velocity comparable to that of NGC~3311 (see Table~\ref{tab:sample}). 
The new deep VEGAS data show that this galaxy, classified as a barred S0, has an extended boxy outskirt,
where we found a prominent arc-like stellar stream on the SW side (Fig.~\ref{fig:NGC3316}, left panel).
From the isophote fit, we built the 2D model of the light distribution, which was subtracted 
from the parent image to obtain the residual map (see the middle and right panels in Fig.~\ref{fig:NGC3316}).
The stellar SW stream clearly stands out from the residuals and, in addition, it seems to be connected
to a bright knot on the N side.
This feature might result from a disrupted dwarf galaxy that interacted with NGC~3316.

To further support this hypothesis,
we derived the integrated magnitudes and colours for the bright knot and the stream,
which are $m_g=19.6 \pm0.2$~mag and $g-r=0.5 \pm 0.3$~mag, and $m_g=21.8 \pm0.2$~mag and $g-r=0.7 \pm 0.3$~mag, respectively.
The $g-r$ colours for both structures are consistent with the range of colours found 
for the dwarf galaxies in
the Hydra I cluster, which is $0.2\leq g-r \leq 1$~mag \citep[][]{LaMarca2022a}.

\subsection{North group and HCG048 group} \label{sec:Ngroup}

\begin{figure*}
    \centering
    \includegraphics[width=1.0\textwidth]{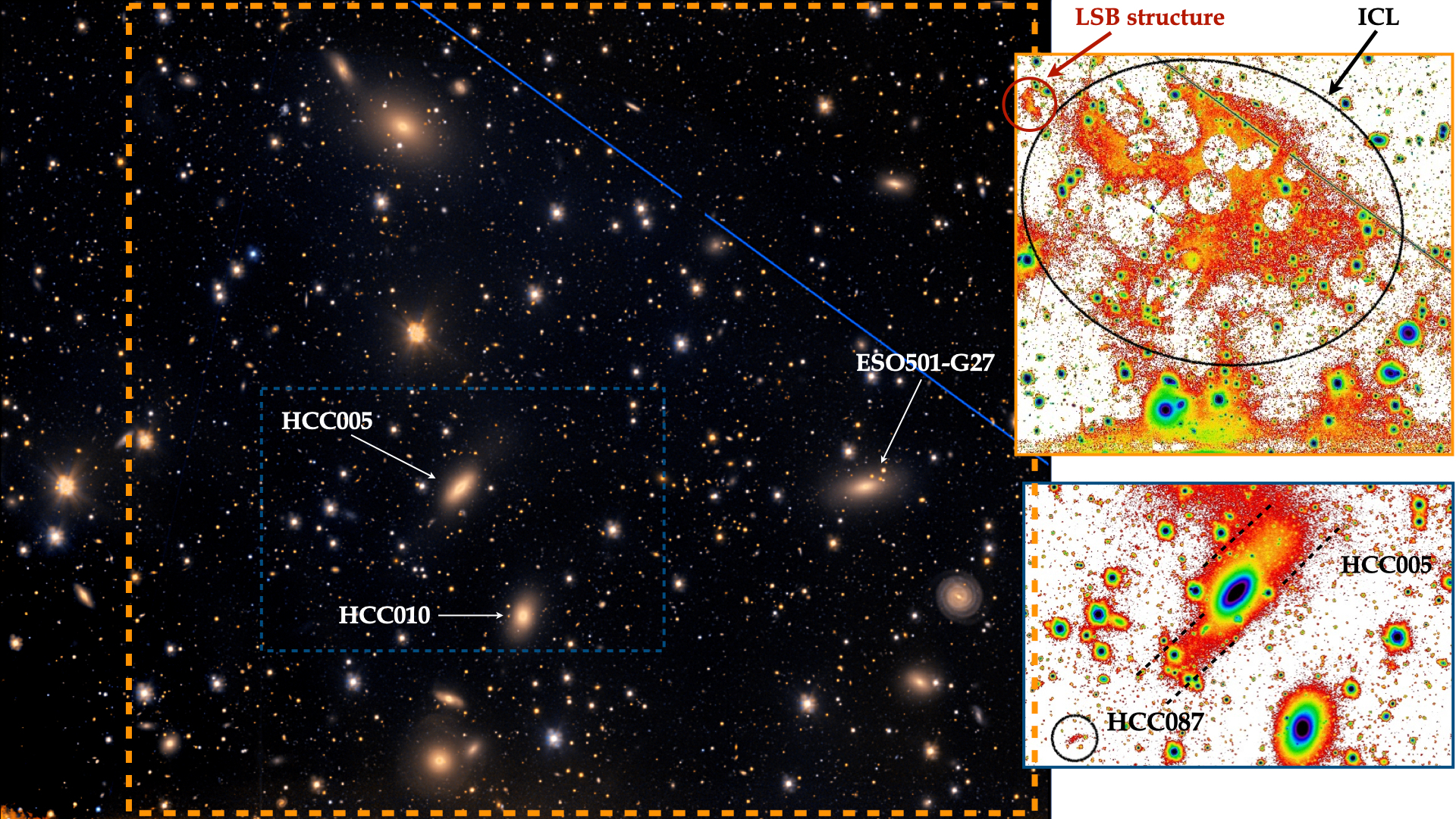}
    \includegraphics[width=9.2cm]{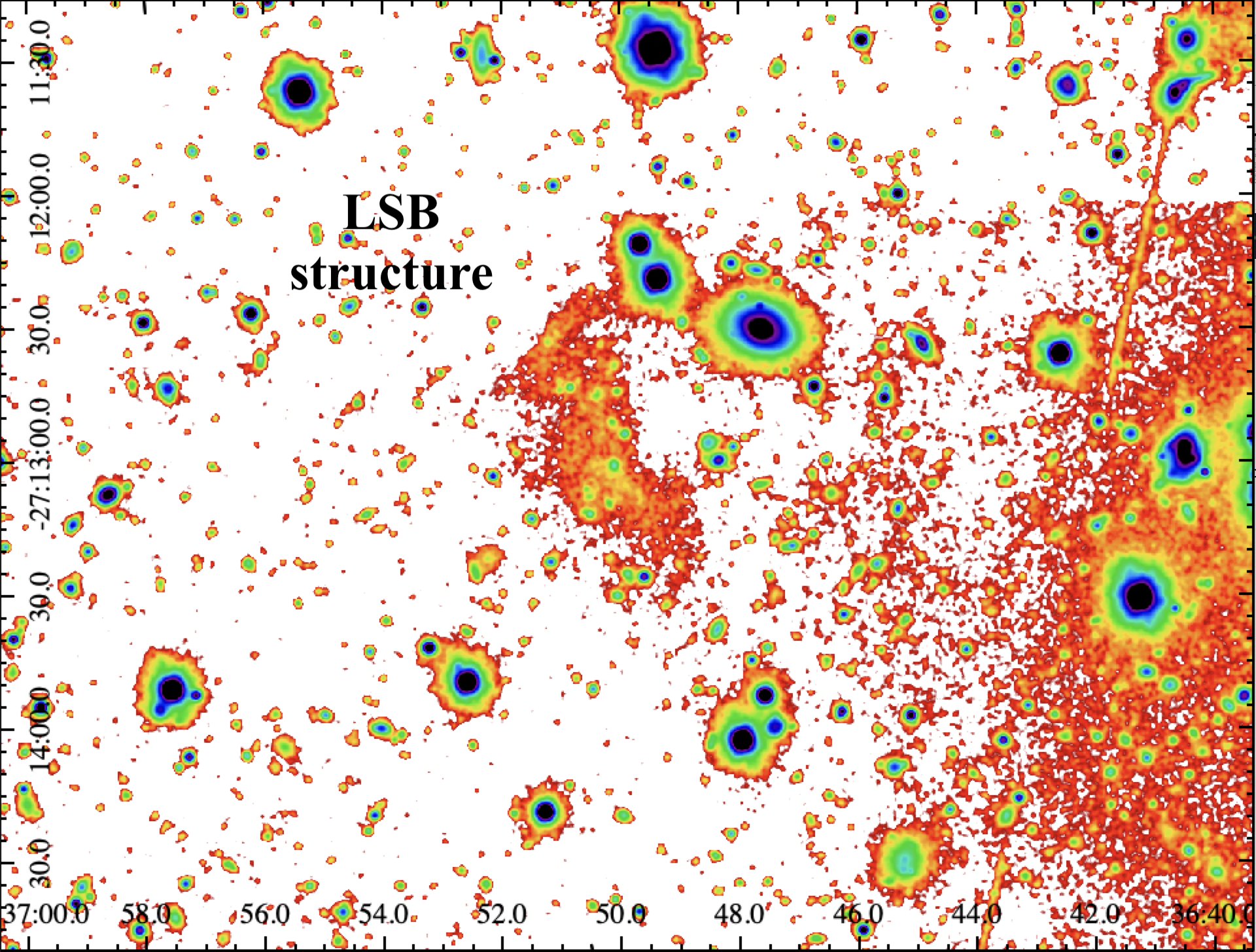}
    \includegraphics[width=9cm]{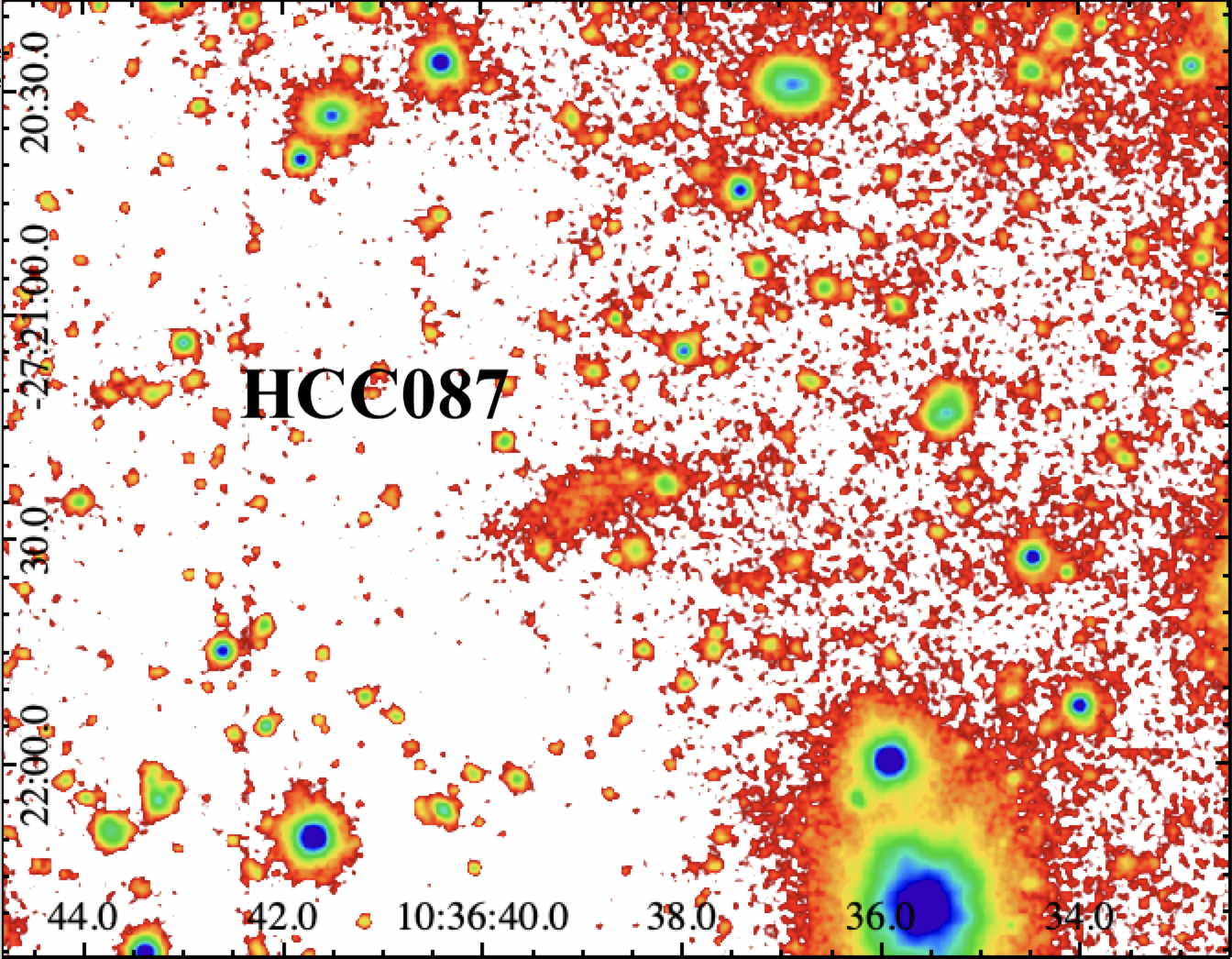}    
    \caption{Structure and LSB feature in the group north of the core of the Hydra I cluster. Top left panel: Colour-composite image of the group, where the brightest members are marked.
    The image size is $15\times15\;arcmin^2$. N is up, and E is left.
    The dashed orange box indicates the region where diffuse light is detected after all bright sources (stars and galaxies) were modelled and subtracted from the image. The residual is shown in the top right panel. On the NE side, the red circle marks the S-shape LSB feature detected on this side of the group, which is also shown in the lower left panel. The dashed blue box marks the region where the peculiar galaxy HCC~005 and the disrupted dwarf galaxy HCC~087 are located, which is shown in the lower right panel. }
    \label{fig:north-group}
\end{figure*}

The over-density of galaxies located in the north with respect to the cluster core is mainly 
distributed along a filament-like structure that extends in projection in the north-south (NS) direction
(Fig.~\ref{fig:north-group}).
In this region of the cluster, we found several LSB features that we describe below.

In the outskirts of the S0 galaxy HCC005, we discovered a thick and extended tail 
in the SE-NW direction that is about twice longer than the inner bright regions of the galaxy (see the middle right panel of Fig.~\ref{fig:north-group}).
This structure has a total integrated magnitude in the $g$ band of $m_g=18.08$~mag 
and an average colour of $g-r=0.73$~mag. The latter value is consistent with the integrated 
$g-r$ colour of the galaxy, which is 0.78 mag (see Table~\ref{tab:sample}), as well as with the $g-r$ colour measured in the outskirts (see colour profile in Fig.~A1).
The similar colours might suggest that this structure is connected with HCC005 as the
result of a recent gravitational interaction.
Alternatively, given the quite regular shape of the outer galaxy isophotes, we cannot exclude that this structure is in projection behind HCC005.

SE of HCC005, we also detect the faint ($\mu_0\sim26.2$~mag/arcsec$^2$, in the $g$ band) 
tidally disrupted dwarf galaxy HCC087, which was previously discovered by \citet{Misgeld2008} 
and was studied in detail by \citet{Koch2012}. It has a peculiar S-shape, which is 
a classic signature of an interaction (see the lower right panel of Fig.~\ref{fig:north-group}).

After we modelled and subtracted all the brightest sources (i.e. galaxies and stars) in this region of the cluster from the parent image, the residual shows an extended patch of diffuse 
light (see the top right panel of Fig.~\ref{fig:north-group}).
Its total luminosity integrated over the whole area is $L \sim 6 \times 10^{9} L_{_\odot}$ and the
integrated colour is $g-r=0.64\pm0.5$~mag. This value is comparable with the average $g-r$ colour
of the stellar envelope around NGC~3311 in the core, which also includes the ICL, where
$0.5 \leq g-r \leq 1$~mag for $R\geq R_{\rm tr}=4.4$~arcmin.

In the same region where ICL is found, we detected a peculiar S-shape LSB feature, located NE 
(see the lower left panel of Fig.~\ref{fig:north-group}). 
It has a total extension of $\sim 2$~arcmin~$\sim 29$~kpc.
The average surface brightness in the $g$ band is $\mu_g\sim 27$~mag/arcsec$^2$, and the
average colour is $g-r\sim0.68$~mag.
The total luminosity is $L_g\sim7 \times 10^7$~L$_{\odot}$.
This structure seems to contain many star-like objects that resemble GCs.
We verified that it is not a high-redshift cluster of galaxies.
Because it is so extended, it seems unlikely that it originated from a disrupted dwarf galaxy.
It might be connected with the ICL in this region of the cluster because the colours are quite similar.
This intruding structure will be the subject of future follow-up studies, in which we focus on
the selection and study of the GC population in the Hydra~I cluster (Mirabile et al., in preparation).

\begin{figure*}
    \centering
    \includegraphics[width=1.\textwidth]{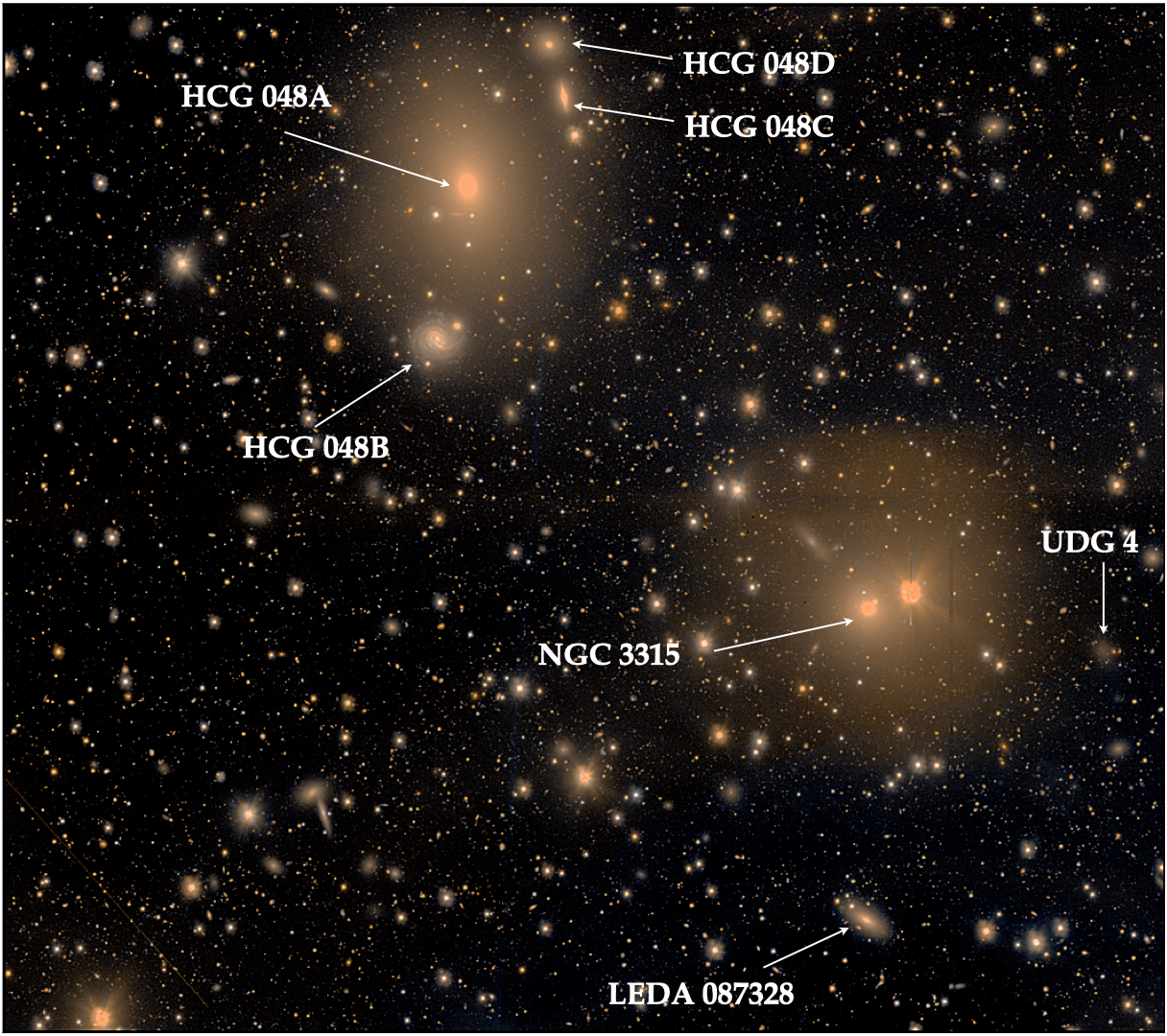}
    \caption{Colour-composite picture of the NE side of the Hydra I cluster. N is up, and E is left. The size of the picture is $18.37 \times 16.21$~arcmin$^2$. 
    In the upper left corner lies the HCG048 group, where the group members are labelled.
    In the lower right corner the other two bright cluster members, NGC~3315 and LEDA~087328 (see also Table~\ref{tab:sample}) are marked. In addition, one of the UDGs detected in the cluster by \citet{Iodice2020}, UDG~4, is also marked in the figure.}
    \label{fig:HCG048}
\end{figure*}

\begin{figure}
    \centering
    \includegraphics[width=9cm]{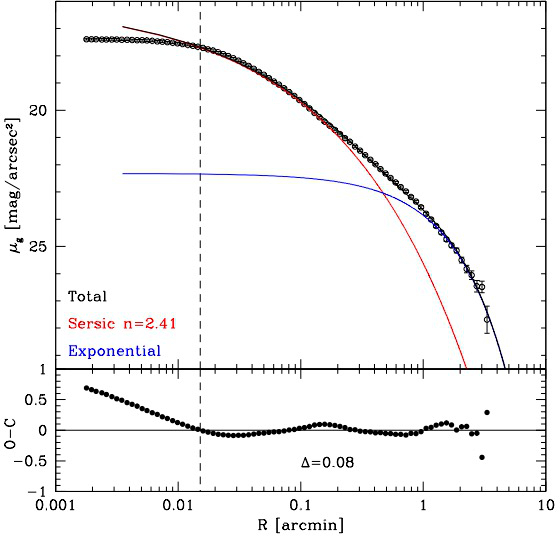}
    \caption{Azimuthally averaged surface brightness profile of HCG48A in the $g$ band. 
    The best 1D multi-component fit is shown in the top panel. 
    In the lower panel, we show the $\Delta$rms residual of the data minus the model (see text for details).}
    \label{fig:HCG048_fit}
\end{figure}

In the north-east (NE) region of the cluster, at about 0.45~Mpc from the core, 
lies the compact group of galaxies HCG~048, which is also associated with the Hydra~I cluster.
This system consists of four galaxies: the brightest member, HCG~048A (IC 2597), an elliptical galaxy, and three
further galaxies that are dimmer and extended, which are the spiral galaxy HCG~048B in the SE and two ETGs, HCG~048C and HCG~048D in the NW (Fig. \ref{fig:HCG048}). Based on the heliocentric velocity, HCG~048B is a foreground galaxy with respect to the other group members (see also Table~\ref{tab:sample}). 
According to \citet{Jones2023}, HCG~048 is likely a false group, and HCG~048C and HCG~048D are a background pair.
The extended stellar envelope associated with the brightest group member, HCG~048A, 
dominates the light distribution of the whole group. 
The surface brightness profile extends out to 3~arcmin~$\sim44$~kpc 
from the centre of HCG~048A and down to $\mu_g \sim 28$~mag/arcsec$^2$ (Fig.~A1).
Using the method described in Sect.~\ref{sec:core}, we performed a multi-component fit to reproduce
these profiles in the $g$ band. The results are shown in Fig.~\ref{fig:HCG048_fit}, and the
best-fit structural parameters are listed in Table~\ref{tab:1Dfit_hcg48a}.
We found that the extended outer stellar envelope is well fitted by an exponential law, 
with a scale length of $r_h=42.29 \pm 0.14$~arcsec, which dominates the light distribution for
$R_{\rm tr}\geq 28.3$~arcmin ($\sim 7$~kpc).
 This component, which includes the stellar halo and the intra-group light, 
has a total luminosity of $L=5.3 \times 10^{10} L_{\odot}$, which corresponds to $\sim 49$\% of the total luminosity of HCG~048A.

\begin{table}[]
\setlength{\tabcolsep}{1.2pt}
\small
\begin{center}
    \caption{Best-fit parameters of the multi-component 1D fit of the surface brightness profiles of HCG~048A in the $g$ band.}
    \begin{tabular}{lccccc}
    \hline
         Law & $\mu_e$ & $r_e$ & $n$ & $\mu_0$ & $r_h$ \\
                   & [mag/arcsec$^2$] & [arcsec] & & [mag/arcsec$^2$] & [arcsec]\\
          (1) & (2) & (3) & (4) & (5) & (6)\\         
         \hline\hline
         Sersic & 20.89$\pm$0.02 & 11.67$\pm$0.12 & 2.41$\pm$0.02 & - & -\\
         Exp & - & - & - & 22.32$\pm$0.01 & 42.29$\pm$0.14\\
         \hline
    \end{tabular}
    \tablefoot{Column 1 reports the empirical law adopted for the multi-component 1D fit.
    Columns 2 to 4 list the effective surface brightness, effective radius, and Sersic's index of the Sersic law for each of the two inner components. Columns (5) and (6) list the central surface brightness and scale length of the exponential component adopted to fit the galaxy envelope.}
    \label{tab:1Dfit_hcg48a}
    \end{center}
\end{table}

\subsection{South-west group} \label{sec:SEgroup}

\begin{figure*}
    \centering
    \includegraphics[width=1.\textwidth]{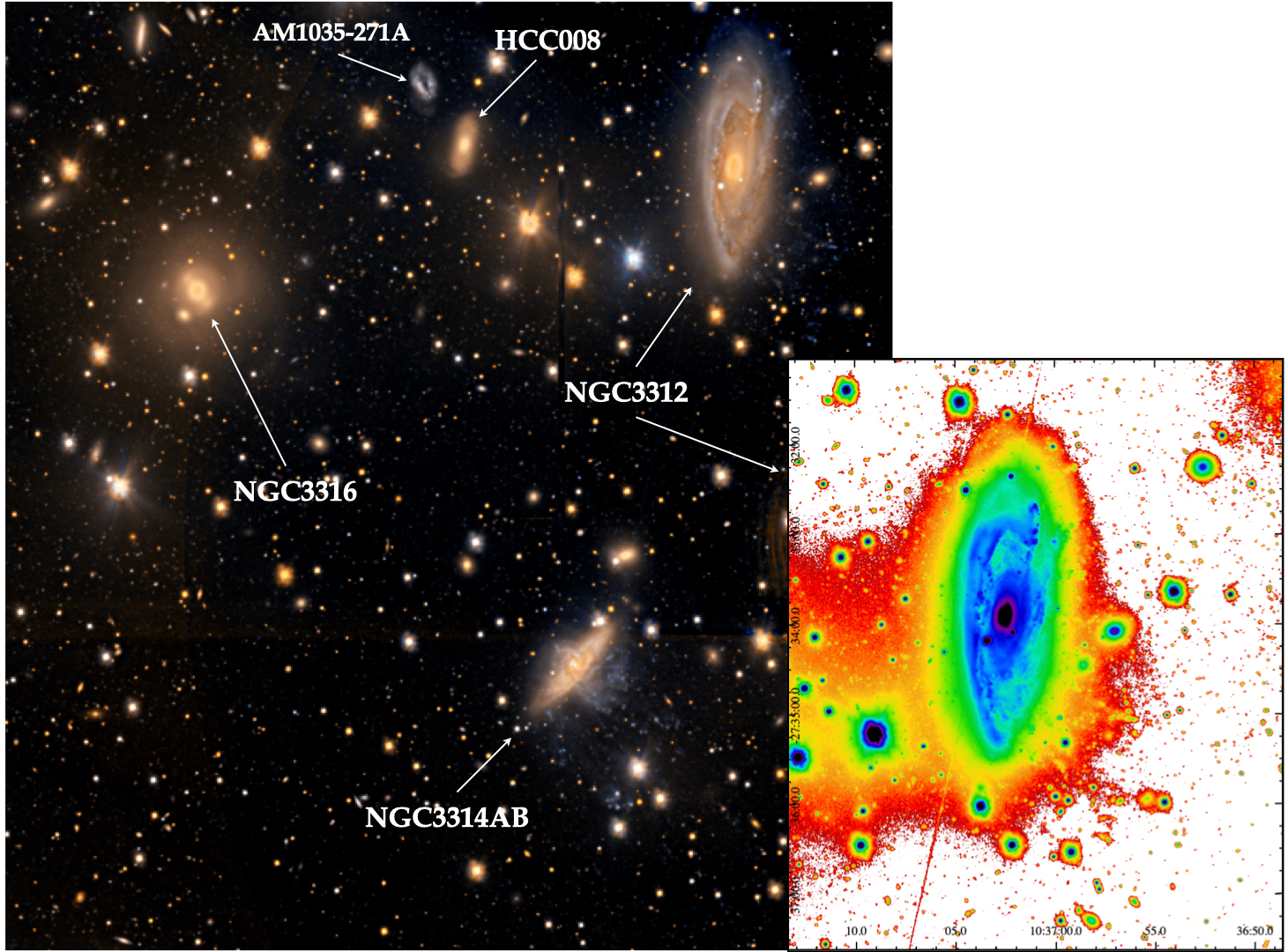}
    \caption{Colour-composite picture of the SE group. N is up, and E is left.
    The size of the image is $12.50\times 13.77\;arcmin^2$. In the bottom part lie the two overlapping spirals, NGC~3314A \textbackslash B \citep[see][]{Iodice2021}, while in the upper right corner lies NGC~3312. In NE, a bright lenticular galaxy is visible, NGC~3316, which is in the background of the group at a similar redshift as the galaxies in the core. The lower right box shows an enlarged region around NGC~3312 to enhance the extended stellar filaments on the SW side of the galaxy, due to RPS. They are marked with the two dashed lines.}
    \label{fig:SE-group}
\end{figure*}

According to the systemic velocities derived by \citet{Christlein2003}, 
a group of foreground galaxies lies in the SE region with respect to the core.
The brightest galaxies of this group are NGC~3312 and NGC~3314A (see Fig.~\ref{fig:SE-group}). 
This group is dominated by late-type galaxies.

\begin{figure*}
    \centering
    \includegraphics[width=1.\textwidth]{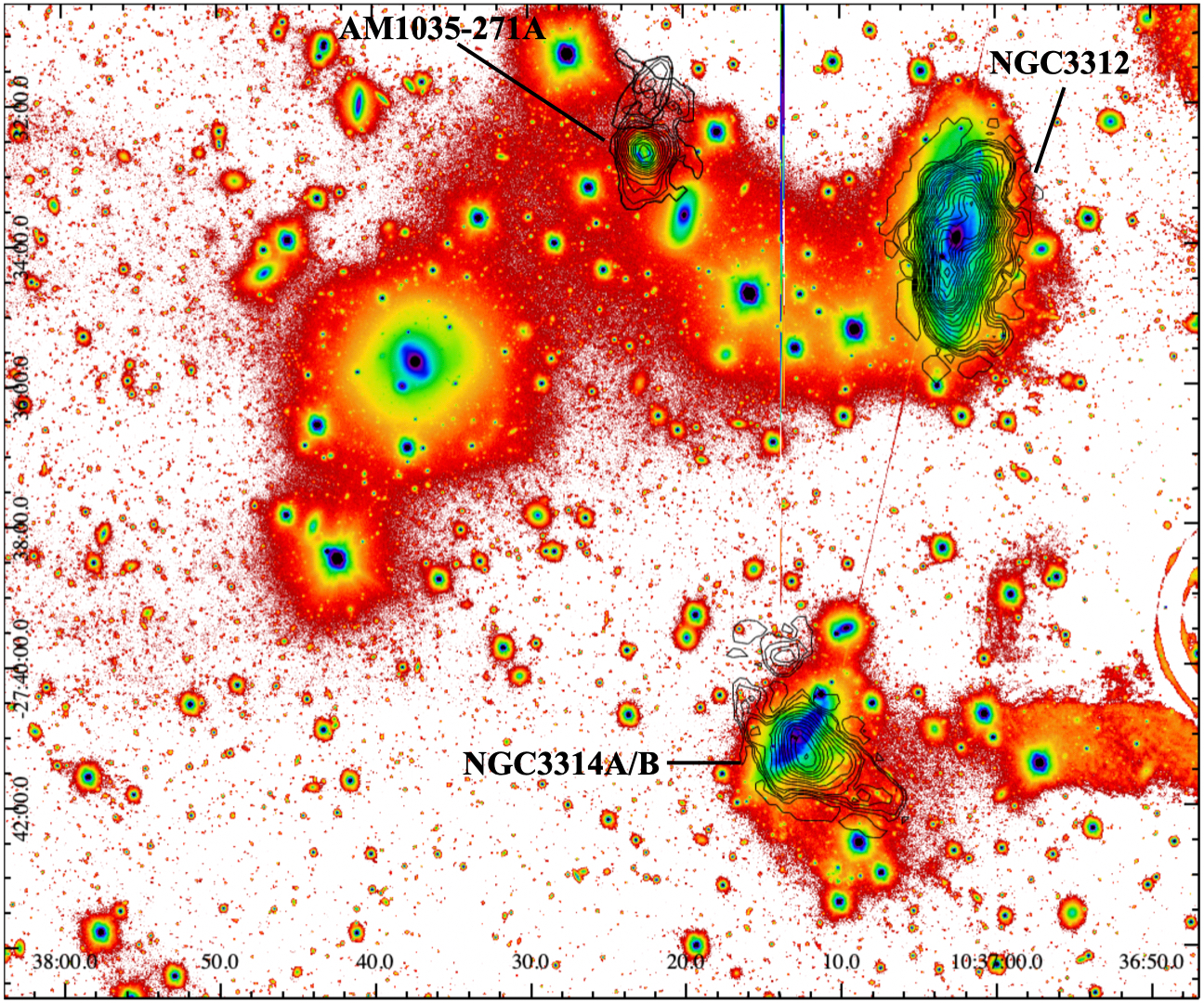}
    \caption{SE group of the Hydra~I cluster of galaxies. The black contours mark the emission of the neutral HI gas from the WALLABY survey \citep{Wang2021}.}
    \label{fig:SE-group_HI}
\end{figure*}

NGC~3312 is a spiral galaxy, the third brightest galaxy inside 0.4~R$_{vir}$ of the cluster.
The deep images presented in this work show stellar filaments extending in the
SW direction (see the right panel of Fig.~\ref{fig:SE-group}). 
This structure, together with the sharp edge on the NE side of the galaxy, 
is consistent with the ongoing RPS process detected in this galaxy based on the HI data
from the WALLABY survey \citep[][]{Wang2021}, and confirmed by the MeerKAT data \citep[][]{Hess2022}. In particular, the stellar filaments found in the VST images are spatially coherent
with those detected from the HI gas (Fig.~\ref{fig:SE-group_HI}). 
The surface brightness profiles for NGC~3312 in the $g$ and $r$ bands and the $g-r$ colour profile from the VST images are shown in Fig.~A1.

South of NGC~3312 lies the system of two spiral galaxies NGC~3314A\textbackslash B, seen in projection 
on top of each other along the line of sight. NGC 3314A is the foreground galaxy, 
with a heliocentric velocity of $cz$=2795~km~s$^{-1}$ \citep[][]{Christlein2003}, 
and NGC~3314B lies in the background with 
$cz$=4665~km~s$^{-1}$ \citep[][]{Keel2001}. 
As a result, the surface brightness and the $g-r$ color profiles, shown in
Fig.~A1, are the contribution to the light of both galaxies, 
which are indistinguishable along the line of sight.
 \citet{Iodice2021} recently studied
NGC~3314A in detail using the deep VST images presented in this work. They found that 
this galaxy shows a network of stellar filaments in the SW direction that are more extended than those 
detected in NGC~3312. They reach a radius of $\sim 50$~kpc from the galaxy centre.
The HI data that are also available for this galaxy (Fig.~\ref{fig:SE-group_HI}) show 
that they have the HI counterpart that can be explained by RPS acting on this galaxy \citep[][]{Wang2021,Hess2022}.

On the E side of NGC~3312 lies a peculiar dwarf galaxy called AM1035-271A 
(see Fig.~\ref{fig:SE-group} and Fig.~\ref{fig:SE-group_HI}), which shows an extended HI emission. 
Being a dwarf, this galaxy is excluded from the sample listed in Table~\ref{tab:sample}.
Having a systemic velocity of $V_{sys}=2763$~km/s \citep[][]{Christlein2003}, this is also
a member of the SE foreground group.
The optical images for this galaxy show a very disturbed morphology, where two luminous
arm-like structures are found on each side with respect to the centre. They might be affected
by dust absorption. 
For this galaxy, differently from the other two giants NGC~3312 and 
NGC~3314A, the HI emission is more extended than the optical light, where HI tails in the
SW direction are found. This indicates that RPS also acts on this galaxy.

\section{Discussion: Assembly history of the Hydra I cluster}\label{sec:discussion}

    
We analysed deep images in the optical $g$ and $r$ bands of the Hydra~I cluster 
of galaxies that were obtained with the VST during the VEGAS project.
The VST mosaic covers  $\sim 0.4$~R$_{vir}$ around the core of the cluster (Fig.~\ref{fig:mosaic}), and we mapped the light distribution down to $\mu_g \sim 28$~mag/arcsec$^2$ on average.

According to the redshift estimates by \citet{Christlein2003}, this region of the cluster 
contains 44 cluster members that are brighter than $m_B \leq 16$~mag. They are listed in Table~\ref{tab:sample}. In addition, more than 200 dwarf galaxies (with $m_B > 16$~mag), 
including UDGs, were recently discovered, and based on the colour-magnitude relation,
they also reside in this area and can be considered cluster members \citep[][]{LaMarca2022a, LaMarca2022b}.
The projected distribution of all cluster members (bright galaxies and dwarfs) suggests
that the bulk of galaxy light is concentrated in the cluster core, as expected, 
where the X-ray emission is also found. In addition, two further galaxy overdensities are found in the north
and SE with respect to the cluster core.

We analysed the light distribution of all bright cluster members in these 
three sub-structures of the cluster. Our results are listed below.
\begin{itemize}

    \item Most of the diffuse light, LSB features, and signs of gravitational interaction between galaxies reside in the core, in the north group, and in the Hickson compact group HCG048.
    \item The extended stellar filaments in the SW direction in the two brightest members 
    NGC~3312 and NGC~3314A, which coincide with the HI emission, and can be associated with the ongoing RPS process.
    \item The light distribution in the core is mapped down to $\mu_g \sim 28$~mag/arcsec$^2$ in the $g$ band and out to $\sim 223$~kpc. We found that the extended and diffuse stellar halo around the BGC, NGC~3311, contributes $\sim 70\%$ of the total luminosity. It therefore is the dominant stellar component in the cluster core. 
    \item The extended stellar envelope around NGC~3311 coincides with the X-ray emission detected in this region of the cluster.
    \item In the cluster core, by comparing the GCs distribution with the surface brightness distribution, we found a shallower profile in the region where the stellar envelope and ICL dominate.
    \item The ICL is detected in the cluster core, north group, and HCG048, with a total luminosity of 
    $L_{\rm ICL}^{core}\sim 1.6 \times 10^{11}$~L$_{\odot}$, $L_{\rm ICL}^{North}\sim 6 \times 10^{9}$~L$_{\odot}$, and $L_{\rm ICL}^{HCG48}\sim 5.3 \times 10^{10}$~L$_{\odot}$, respectively.
   {\item  The specific frequency of GCs increases in the ICL region, and there is a deficit of dwarf galaxies in the very centre.}

\end{itemize}

The presence of sub-structures, the detection of the stellar debris as typical features of the 
gravitational interactions, and the ongoing RPS further confirm that the Hydra~I cluster 
is still in an active assembly physe. By combining the projected phase-space (PPS) of
the Hydra I cluster with the observed properties, including the new measurements and detection 
reported in this paper, we traced the different stages of the assembly history.

\begin{figure}
   \centering
    \includegraphics[width=12cm]{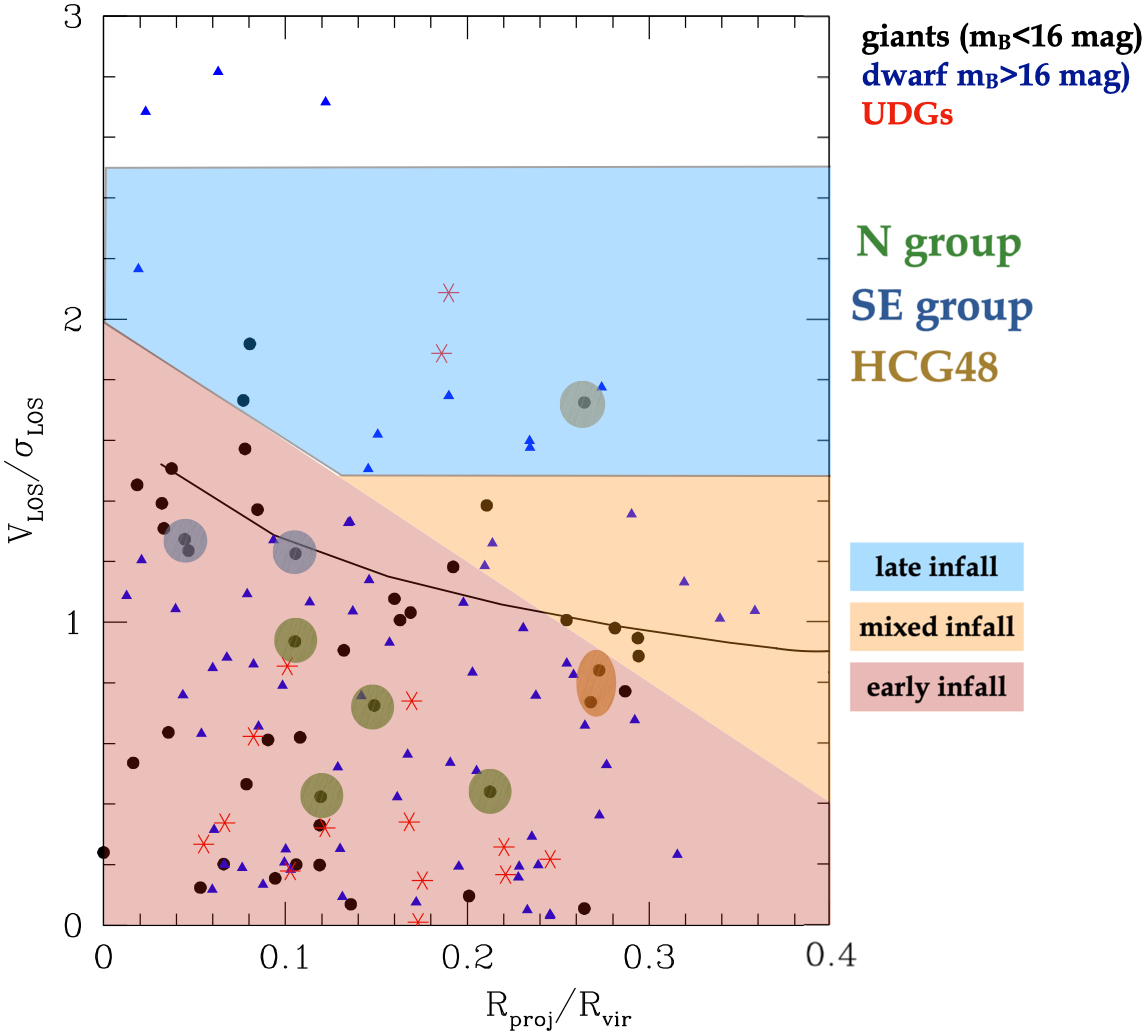}
    \caption{Projected phase-space diagram of the cluster members in Hydra~I inside the VST mosaic, i.e. $\sim0.4R_{\rm vir}$. The giants and brightest galaxies ($m_B <16 mag $) and the dwarf galaxy population by \citet{Christlein2003} are marked with filled black circles and blue triangles, respectively. 
    The newly confirmed UDGs in the cluster by \citet{Iodice2023} are marked with red asterisks.
    The solid black line corresponds to the cluster escape velocity.
    The shaded regions of the diagram indicate the infall times as predicted by \citet{Rhee2017}, and reported in the legend on the right. The brightest members of the sub-groups of galaxies identified in the clusters (i.e. north group, SE group, and HCG48 group) are also marked with coloured circles, as listed on the right side of the figure.}
    \label{fig:PPS}
\end{figure}

The PPS was derived by using the redshift for all cluster members published by \citet{Christlein2003} 
and those recently measured for the LSB galaxies by \citet{Iodice2023}. 
The PPS is shown in  Fig.~\ref{fig:PPS}.
To derive the $V_{LOS}/\sigma_{LOS}$ for the PPS, we assumed  
$V_{\rm sys} = 3683 \pm 46$~km/s and $\sigma_{\rm cluster}\sim700$~km/s as 
the mean cluster velocity and velocity dispersion, respectively  
\citep{Christlein2003,Lima-dias2021}.
Based on the PPS, inside $0.4R_{\rm vir}$ of the cluster (which is the region studied in this work)
most of the cluster members are located in the early-infall region, 
where galaxies joined the cluster potential 
more than 10 Gyr ago \citep[][]{Rhee2018}.
Therefore, they have experienced repeated interactions and merging, which contributed to the
build-up of the stellar haloes and ICL. This is indeed the region where we have detected the
extended stellar halo and ICL around the BGC NGC~3311 in the core and the 
diffuse light in the north group. In both sub-structures of the cluster, we also found many
signs of past interactions between galaxies in the outskirts 
in the form of tidal tails or stellar streams or ripples (Fig.~\ref{fig:core_residual}). PPS has also been used by \citet{Forbes2023} to examine a sample of UDGs in the Hydra I cluster. This analysis led to the conclusion that UDGs are among the earliest infallers in Hydra.  

In the PPS, the SE group and HCG048 group are found close to the escape velocity of the cluster, 
except for the spiral galaxy HCG48B, which is located in the late infall region.
The deep images presented in this work have further confirmed that the SE group, falling into
the cluster potential, interacts with the hot intra-cluster medium, inducing the RPS, as seen in the HI emission and from the optical counterparts we showed here (Fig.~\ref{fig:SE-group_HI}).

Based on the PPS, the ICL is detected in the virialised regions of the cluster, that is,
the core and north group, where signs of tidal interactions and accretion of small satellites
are still present. This might suggest that these are the main channels that contribute
to the ICL formation. These processes create the population of  intra-cluster GCs,
which are also detected in the cluster core, having a cospatial distribution with the diffuse
stellar envelope around NGC~3311.
In addition, inside $\sim 15$~arcmin
from the centre of NGC~3311 ($\sim 223$~kpc), where the extended diffuse stellar halo 
and X-ray emissions dominate, we found a decreasing projected number density 
of dwarf galaxies  (Fig.~\ref{fig:Dwarf_ND}). This is an indication that these 
low-mass systems might have been disrupted by the strong tides in this region of the cluster,
contributing to the diffuse stellar component around NGC~3311.

Based on the PPS, the small compact group of HCG048 located NE from the core will be accreted onto the cluster. The intra-group light is clearly visible around the dominant group member (Fig.~\ref{fig:HCG048}),
HCG048A, and it will contribute to the total budget of the ICL when the process is completed.
As described in Sect.~\ref{sec:intro}, this would further support the idea that the 
pre-processing in groups is one of the formation channels of the ICL. 

The total amount of the ICL, estimated on the whole region of the cluster covered by the VST mosaic, 
is $L_{\rm ICL} \sim 2.2 \times 10^{11} L_{\odot}$. 
This includes the ICL in the core, the ICL in the north, and the intra-group light in HCG048.
The total luminosity of the cluster inside 0.4 $R_{\rm vir}$ that is covered by the VST mosaic is
$L_{TOT} \sim 1.8 \times 10^{12} L_{\odot}$, and the fraction of the ICL therefore is
$f_{ICL}= L_{\rm ICL}/L_{TOT}\sim 12\%$.
This value is fully consistent with values of $f_{ICL}$ for other clusters of galaxies with 
a virial mass comparable to that of Hydra I, that is, M$_{vir}\sim 10^{14}$~M$_\odot$.
In particular, for clusters and groups in the nearby Universe (i.e. $z\leq 0.05$), 
in the mass range $13.5 \leq log $~M$_{vir} \leq 14.5$~M$_\odot$, the ICL fraction
ranges from $\sim10\%$ up to $\sim 45\%$ \citep[see][and references therein]{Kluge2020,Ragusa2023}.

In summary, the large covered area and the long integration time that were available for mapping the central 
regions of the Hydra I cluster down to the LSB regime allowed us to trace the final 
stage of the mass assembly inside the core of the cluster. In this region, the ICL is already in place, 
but many remnants of the interactions and mergers between galaxies are still present, 
and they also contribute to the total amount of diffuse light.
We have acquired deep VST images for three additional fields around the core of the cluster, 
covering the E and SE regions. For future work, we aim to explore these lower-density regions to derive an extended density map of the cluster members and
compare their structure and colours with those found in the core. This analysis would enrich 
the picture we have traced of the assembly history of the Hydra I cluster.

Finally, the region of the Hydra I cluster will be covered by the ongoing Euclid Wide Survey
(Mellier et al. 2024) with the Euclid space telescope and the approaching Legacy Survey of 
Space and Time (LSST) with the Rubin Observatory \citep[][]{Brough2024}. 
The data presented in this paper approach in depth and resolution 
those expected from the 
 observing facilities cited above, even if in limited areas of the sky. This work 
(as many of the focused deep multi-band imaging surveys that were carried out in the last decade) can therefore
provide a preview of the science that will soon be delivered by the new surveys, and it represents a
testbed for building the knowledge required for managing the upcoming massive data sets.

\begin{acknowledgements}
The authors are very grateful to the referee, Emanuele Contini, for his comments and suggestions which helped to improve and clarify the work.
This work is based on visitor mode observations collected at the European Southern Observatory (ESO) La Silla Paranal Observatory within the VST Guaranteed Time Observations, Programme ID: 099.B-0560(A). EI, MS and MC acknowledge the support by the
Italian Ministry for Education University and Research (MIUR) grant PRIN 2022 2022383WFT “SUNRISE”, CUP C53D23000850006 and by VST funds.
EI, MS, MC  acknowledge funding from the INAF through the large grant PRIN 12-2022 "INAF-EDGE" (PI L. Hunt).
ALM acknowledges financial support from the INAF-OAC funds. EMC acknowledges support by Padua University grants DOR 2020-2023, by Italian Ministry for Education University and Research (MIUR) grant PRIN 2017 20173ML3WW-001, and by Italian National Institute of Astrophysics (INAF) through grant PRIN 2022 C53D23000850006. GD acknowledges support by UKRI-STFC grants: ST/T003081/1 and ST/X001857/1. 
Authors acknowledge financial support from the VST INAF funds.


\end{acknowledgements}

 \bibliographystyle{aa.bst}
  \bibliography{Hydra}
  
\begin{appendix}

\section{Surface brightness profiles}\label{sec:sb_prof}
This appendix, available on Zenodo
(\url{https://zenodo.org/records/13123011}), provides the azimuthally averaged surface brightness and
colour profiles of the sample galaxies listed in
Table~\ref{tab:sample}. 

\end{appendix}
\end{document}